# Lessons Learned from Photovoltaic Auctions in Germany

Taimyra Batz Liñeiro*[1] and Felix Müsgens*

*Brandenburg University of Technology Cottbus-Senftenberg (BTU)

**Abstract:** Auctions have become the primary instrument for promoting renewable energy around the world. However, the data published on such auctions are typically limited to aggregated information (e.g., total awarded capacity, average payments). These data constraints hinder the evaluation of realisation rates and other relevant auction dynamics. In this study, we present an algorithm to overcome these data limitations in German renewable energy auction programme by combining publicly available information from four different databases. We apply it to the German solar auction programme and evaluate auctions using quantitative methods. We calculate realisation rates and—using correlation and regression analysis—explore the impact of PV module prices, competition, and project and developer characteristics on project realisation and bid values. Our results confirm that the German auctions were effective. We also found that project realisation took, on average, 1.5 years (with 28% of projects finished late and incurring a financial penalty), nearly half of projects changed location before completion (again, incurring a financial penalty) and small and inexperienced developers could successfully participate in auctions.

*Keywords:* Renewable energy; Auction; Solar power; Photovoltaic; Germany

## 1 INTRODUCTION

Auctions have become the primary instrument for promoting renewable energy around the world (IRENA, 2017; Sach et al., 2018). As of the end of 2020, more than 145 GW of renewable capacity has been awarded via auctions (IEA, 2020). Alongside the rise of auctions to promote RES expansion, research on this topic has also increased. The literature focuses on optimising auction design to successfully meet policy objectives; however, nearly all empirical articles measure auctions' success based only on the awarded capacity and the resulting average bids. These measures, however, are insufficient to answer important questions about the auctions' effectiveness, bidding behaviour and appropriateness of the auction design. The main reason behind this dearth of empirical research is the fact that it is difficult to derive quantitative results from the limited data published on auction results.

---

[1] Corresponding author: Taimyra Batz Liñeiro, BTU Cottbus-Senftenberg, Siemens-Halske-Ring 13, 03046, Cottbus, Germany. E-mail: Taimyra.BatzLineiro@b-tu.de.

One of our key contributions is an innovative approach entailing the integration of information from various databases into a body of detailed, project-specific information on auction results pertaining to the German PV auction programme. The project-specific characterisation enables us to provide quantitative answers to several interesting research questions. We estimate realisation rates and provide explanations for the marked variation between the early and recent auctions, enabling insights into expected future realisation rates. Furthermore, our work quantifies project construction duration, changes in location, and bid height. The first is vital for regulators designing auction programs, as they face a conflict between adding installed capacity as soon as possible and giving investors time to enter an auction without excessive upfront payments. The second is interesting because it involves a penalty, effectively reducing the net subsidies paid out by the programme. The third allows for a much more elaborate assessment of competition and bidder behaviour.

In addition, we test for common hypotheses in the literature regarding the ability of large and experienced developers to i) drive smaller agents out of the market and ii) derive larger profits by more accurately estimating the marginal bid under the pay-as-bid scheme. Additionally, we test whether a lack of experience increases the risk of underbidding (winner's curse) or results in project delays. Finally, we propose a regression model to compare the effects of competition and PV technology costs on single bids.

The remainder of this paper is organised as follows. In Section 2, we examine the literature on auction evaluation and detail the design elements and results of German PV auctions. In Section 3 we describe our hypothesis framework, data sources and data-processing methodology. In Section 4, we present and discuss our results. Finally, in Section 5, we conclude with lessons that can be drawn from the German experience.

## 2 THEORETICAL BACKGROUND

### 2.1 Literature

Since 2000, 53 peer-reviewed research articles addressing the topic of auctions promoting RES expansion have been published.[2] Several desktop reports have also been released (IEA, 2020; IRENA, 2018, 2017, 2013; REN21, 2018; Tiedemann et al., 2019; USAID, 2016). The literature largely centres around the study of auction design (Anatolitis & Welisch, 2017; Black, 2005; Del Río & Mir-Artigues, 2019; Eberhard & Kåberger, 2016; Gephart et al., 2017; Kreiss et al., 2017; Kruger & Eberhard, 2018; Mora et al., 2017; Shrimali et al., 2016; Voss & Madlener, 2017; Welisch & Kreiss, 2019).

---

[2] Scopus search (TITLE (auction OR tender) AND TITLE (renewable OR solar OR wind OR PV OR photovoltaic)) AND (LIMIT-TO (DOCTYPE, 'ar') OR LIMIT-TO (DOCTYPE, 'cp')) resulted in 76 articles. Nineteen addressed auctions in the spot market context; therefore, they were removed from the set of literature.

This study, however, focuses on two aggregated measures: i) total awarded capacity and ii) final capacity-weighted average bid. While the literature explicitly highlights the importance of effectiveness (i.e., how much capacity is ultimately built; (del Río, 2017; del Río et al., 2015; Hochberg and Poudineh, 2018; Mora et al., 2017; Toke, 2015), few articles quantify project realisation rates (Batz and Müsgens, 2019; Bayer et al., 2018a; Gephart et al., 2017; Shrimali et al., 2016; Winkler et al., 2018). This discrepancy largely stems from the lack of appropriate auction data. First, the quantification of realisation rates requires waiting for construction deadlines to be met. Second, once a deadline is met, information about the project's construction status is often difficult to access or simply unavailable (Bayer et al., 2018b; Cassetta et al., 2017; Shrimali et al., 2016; Tiedemann et al., 2016; Winkler et al., 2018).

This lack of information also prevents the assessment of other interesting auction dynamics, such as the effects of competition, geographical distribution and developer size on bids and effective deployment. Usually, assertions about these effects are purely theoretical (del Río et al., 2015; del Río and Mir-Artigues, 2019; Gephart et al., 2017; Hochberg and Poudineh, 2018; IRENA, 2017; Kruger and Eberhard, 2018; Müsgens and Riepin, 2018). To the best of our knowledge, only two research articles have empirically assessed these effects. However, the results of these research papers are weakened by the limited reliability of their data in terms of both quantity and quality. Bayer et al. (2018b) compare auction results from Brazil, France, Italy and South Africa by analysing bids, realisation rates, project duration and market concentration. However, they warn that complete information about the owners is only available for Brazil—the estimations for the other countries are based on a limited sample of publicly owned projects. Cassetta et al. (2017) use data from the Italian wind auction programme and regress all bids (awarded and unawarded) onto proxies for developer size, developer experience, level of competition and some additional project characteristics. However, they did not analyse realisation rates due to a lack of information.

Finally, the literature addresses three hypotheses on the advantages of large developers (Bayer et al., 2018b; Cassetta et al., 2017; del Río et al., 2015; del Río and Mir-Artigues, 2019; Gephart et al., 2017; Hochberg and Poudineh, 2018; IRENA, 2017; Müsgens and Riepin, 2018). First, large developers can make use of cost advantages to bid systematically lower than small developers, pushing them out of the market. Second, large developers can make stronger marginal bid estimates than small developers, leading to higher profits. Third, small and inexperienced developers struggle to estimate their costs and manage their projects, leading to underbidding and project delays. However, once again, it must be noted that the existing literature on RES auctions is largely normative on these issues.

Our research adds to the existing empirical analyses by conducting a deep assessment of the PV auction programme in Germany. We estimate each auction's project realisation rate, project duration, and single bid values. Additionally, we test whether project duration is affected by location or location changes. Furthermore, we analyse the degree to which the German PV auction design favours large and experienced actors and evaluate the role of competition amid falling technology costs.

## 2.2 The German PV Auction Design

Auctions for RES were introduced in Germany in 2015 as a part of a pilot programme stemming from the Solar Large-Scale Tender Regulation (*Freiflächenausschreibungsverordnung* [FFAV]). The government aimed to increase the installed capacity from ground-mounted PV by 400 MW p.a. (FFAV, 2015, para. 1). The pilot was considered a success; demand was high, and subsidy levels decreased by nearly 25%. As a result, since 2017, all solar installations with a size equal to or higher than 750 kW have been required to take part in an auction to be granted a subsidy. With the introduction of the RES Act 2017, the auction procedure was regulated in further detail; these regulations now included general conditions for bidders, penalties, tariff reductions and rules about information disclosure. A summary of the most important regulatory provisions can be found in Table 1; Appendix A provides more detailed information on this subject.

The main objective of the German auction design is to guarantee PV deployment at competitive subsidy levels. To that end, the design incorporates a ceiling price, financial pre-qualifications and penalties for non-compliance. Previous international experience with RES auctions, however, suggests that penalties may be insufficient incentives (Bayer et al., 2018a). Given the low penalties for non-compliance (7% of total project costs[3]), German auctions are essentially, by design, set up as 'options to build' (Müsgens and Riepin, 2018).

Furthermore, the limits on project size indicate secondary objectives, such as developer size and diversity. While larger projects may benefit from economies of scale (static efficiency), small installations have sometimes been shown to improve dynamic efficiency due to increased competition (del Río et al., 2015; Fraunhofer Ise et al., 2014). However, as developers are not limited in the number of auctions in which they participate or the number of projects on which they bid in each auction, the size limit has relatively small implications.

**Table 1: General Auction Design and Regulation**

| Dimension | Description |
|---|---|
| Introduction | 2015 |
| Regulation | Ground-mounted PV tendering regulation (pilot, 2015–2016) <br> RES Act 2017 (2017 onwards) |

---

[3] Calculation based on average installation cost of 700 €/kW for large-scale PV installations (Fraunhofer ISE, 2018).

| | |
|---|---|
| Auction product | Capacity (MW) |
| Pricing rules | Pay-as-bid scheme (except 2nd and 3rd auctions, which use a uniform-price scheme) |
| Frequency | Quarterly (see Appendix B for more details) |
| Ceiling price | Yes |
| Bid size | 750–10,000 kW |
| Financial pre-qualification | The bid bond amounts to 50 €/kW. An initial security deposit of 5 €/kW must be paid when a bid is made; an additional security of 45 €/kW must be paid by the tenth working day following the public announcement of the award. In line with §37a of the RES Act 2017, the second payment can be reduced to 20 €/kW if full land-use documentation is provided alongside the bid. |
| Penalty for non-compliance | If the second security is not paid, the penalty equals the first security (i.e., 5 €/kW). If 5% or more of the bid is cancelled (not built within a 24-month period), the penalty amounts to 50 €/kW of the cancelled capacity. The penalty can be reduced to 25 €/kW if the second security was reduced through the provision of full land-use documentation. |
| Tariff reduction for missing the first deadline | There is a tariff reduction of 0.3 €/kWh if the units are not commissioned within 18 months of the winner being notified. If a bid has been subdivided into several projects, the reduction applies only to the projects that have been delayed. |
| Tariff reduction for changing project location | There is a tariff reduction of 0.3 €/kWh if the final construction location does not, at least in part, conform to the location indicated in the bid. Reductions are additive. As a result, missing the first deadline and changing project location leads to a 0.6 €/kWh reduction. |
| Form of support auctioned | Sliding FIP (also referred to as a one-sided CfD) |
| Support duration | 20 years |

## 2.3 Auction Results Published by BNetzA

Since 2015, the German Federal Network Agency (*Bundesnetzagentur* [BNetzA]) has carried out 23[4] auctions; another 13 are planned through the end of 2022 (see Appendix B). As stipulated by §35 RES Act 2017, BNetzA publishes results after each auction is held, including the bid ID[5], the developer's name, the number of projects comprising each bid, the planned construction site and aggregated information about the auction (e.g., total capacity tendered and awarded, number of submitted and awarded bids, maximum and minimum awarded bids, capacity-weighted average bid value). With a significant time lag[6], BNetzA also publishes aggregated realisation rates per auction.

Key results published by BNetzA (see Figure 1) show that—except for the first auction, when tendered capacity was extraordinarily high—demand for the programme has always surpassed tendered capacity. Bids decreased by about 53% between the first auction in April 2015 and the tenth auction—which was the lowest point—in February 2018. However, since 2018, there has been a fairly neutral, slightly upward trend in the average bid. Interestingly, an increase in the bidding range often corresponds to increases in both the average bid and auction volumes. Higher volumes may attract a wider range of bidders with different bidding strategies. At the same time, the stabilization of PV prices makes it more difficult for developers to anticipate price fluctuations.

---

[4] As of 15/11/2020.
[5] The bid ID is an 11-character code indicating its auction year, round, and placement. For example, FFA15-1/129 is the code for the winning bid from the first auction round in 2015. In order of arrival, this bid was the 129th of the 170 bids placed during that round.
[6] As of 15/11/2020, auction data is only available until the end of 2017.

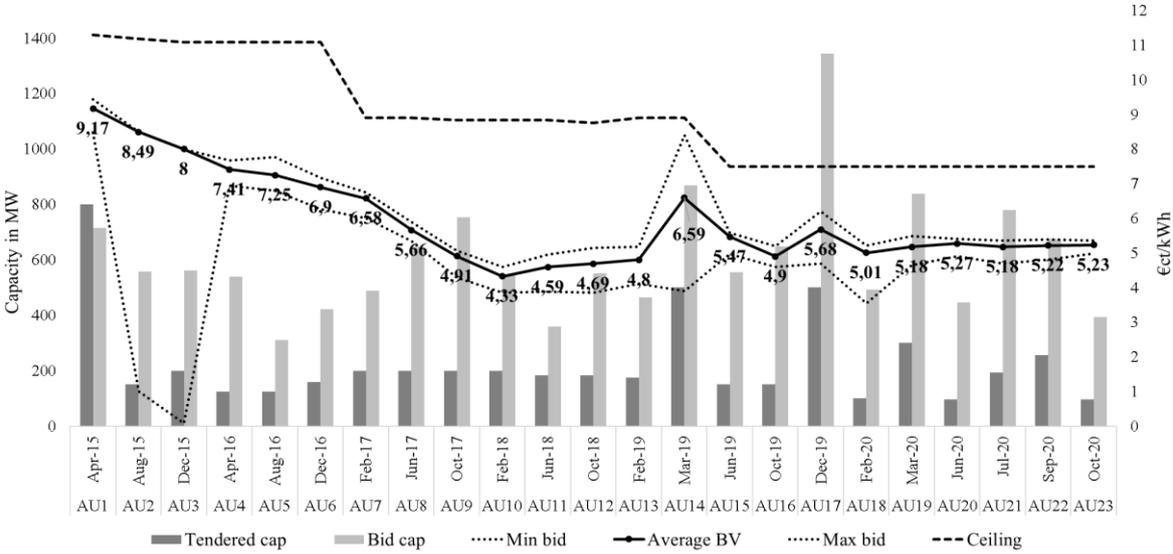

**Figure 1: Auction Results**

While these findings are already quite insightful (e.g., it confirms the increasing competitiveness of PV projects), there is still a lot of missing information. First and foremost, BNetzA provides no information regarding individual bids. Additionally, it does not publish information about the development status of projects at the end of the 18-month (deadline without tariff reduction) or 24-month (deadline with tariff reduction) construction periods. Moreover, it discloses no information on location changes. As both delays and location changes lead to *additive* penalties of 0.3 €/kWh, this information is needed to quantify the net revenues received by PV projects, which more accurately reflect the true competitiveness of PV installations and the true costs of these projects for consumers.

While BNetzA publishes information on final realisation rates, they do so with a substantial delay. Hence, a comparison of auctions with high and low subsidy levels is not possible using BNetzA data alone. Furthermore, little is known about the pace of realisation and its underlying dynamics.

# 3  METHODOLOGY AND DATA

## 3.1 Lead Questions

The set of questions and hypotheses this research aims to solve are summarised in Table 2.

**Table 2: Lead Questions Answered in Our Analysis.**

| Question | Measure/Hypothesis/Method | Data Requirements |
| --- | --- | --- |
| 1. Are projects being realised? | | |
| | Realisation rates | · Project status |

| Question | Measure/Hypothesis/Method | Data Requirements |
|---|---|---|
| | | · Project size |
| 2. How long do German developers take to construct PV projects? | | |
| | Duration (under the assumption that project construction begins immediately after the first announcement of winners). Occurrence of realisation penalties | · Winner's announcement date<br>· Project's commissioning date |
| 3. What percentage of projects change location? | | |
| | Occurrence of location change | · Location at bid submission<br>· Location at commissioning |
| 4. What is the difference between the net bid values (i.e., including penalties for delays and location changes) and the full bid values (i.e., the actual bids)? | | |
| | Net bid values<br>Full bid values | · Market premia<br>· Market values<br>· Occurrence of location change<br>· Occurrence of post-deadline realisation |
| 5. Do geographical aspects affect bids and construction periods? | | |
| | $H0_{5.1}$: There is no difference in average project duration between projects with and without location change<br>$H0_{5.2}$: There is no difference in average subsidy level between projects with and without location change<br>$H0_{5.3}$: There is no difference in average project duration between projects built in the north and those built in the south of Germany<br>$H0_{5.4}$: There is no difference in average subsidy level between projects built in the north and those built in the south of Germany<br>Method: Mann–Whitney–Wilcoxon test | · Occurrence of location change<br>· Regional location<br>· Project duration<br>· Full bid values |
| 6. Does the auction design favour large players? | | |
| Smaller developers are pushed out of the market<br><br>Large developers have an advantage to win higher profits<br><br>A lack of experience may result in underbidding or project delays | $H0_{6.1}$: The proportion of new developers decreases steadily over time<br>$H0_{6.2}$: The proportion of small developers among awarded bids decreases over time<br>$H0_{6.3}$: There is no correlation between developer size (capacity) and accuracy at estimating the marginal bid<br>$H0_{6.4}$: There is no difference in average project duration between experienced and inexperienced developers<br>$H0_{6.5}$: There is no difference in average subsidy level between projects realised by experienced developers and those realised by inexperienced developers<br>Methods: Proportion of new developers over time, proportion of small, awarded developers over time, Correlation test, Mann–Whitney–Wilcoxon test | · Aggregated developers<br>· Small developers are defined as those with a maximum of 2 MW of won capacity.<br>· Developer size as accumulated capacity won throughout the programme.<br>· Developer experience<br>· Maximum bid per auction.<br>· Project duration<br>· Full bid values |
| 7. Does competition influence bid values? | | |
| | $H0_{7.1}$: Competition has a significant negative effect on bid values<br>$H0_{7.2}$: The reduction in technology costs has been a major factor behind the auction results<br>Method: Regression analysis | · Competition as the ratio of bid capacity to tendered capacity.<br>· Technology costs (PV cost index). |

## 3.2 Methodology and Data

Answering our research questions requires data that are neither readily accessible nor commercially available. We overcome data limitations to produce detailed auction results by combining five different databases published by two organisations. Two of the databases are published by BNetzA. The first source comprises the auction results introduced in Section 2.3. The second source is a unit register (*Marktstammdatenregister*) containing all RES and conventional units that have ever been registered as energy generators in Germany. This register is regularly updated and contains specific unit information, including the unit's ID and associated technology, capacity, location and developer.

The other three databases are published by the German transmission system operators (TSOs; Amprion, 50Hertz, Transnet BW and Tennet). The third source is the payment register[7] (*Bewegungsdaten*), it is is published at the end of each year with a one-year lag and includes: unit ID, RES Act subsidy tariff ID, generation and payment. The fourth source corresponds to the monthly market values for solar power,[8] which are an integral component of the calculation of bid values (see eq. (7) below). Market values are updated monthly with a one-month lag. Finally, the fifth source is the RES Act tariff register, which includes tariff IDs, descriptions, and values and is updated each year without a lag. An overview of our data and the unit-identification process is depicted in Figure 2.

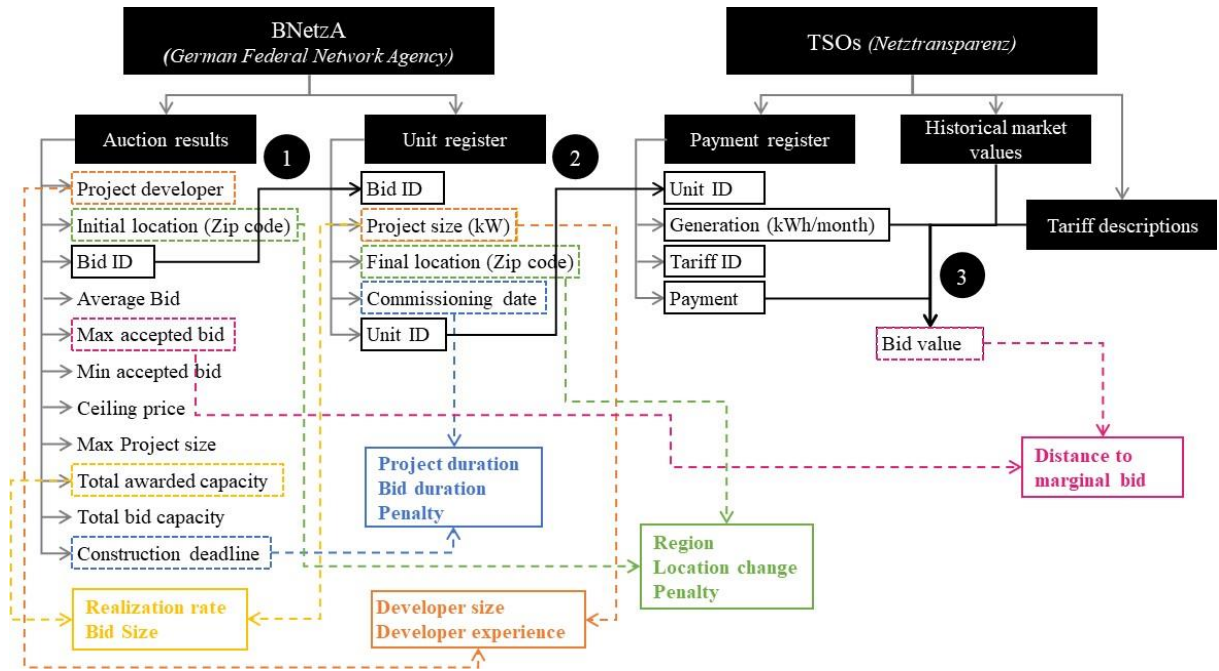

**Figure 2: Data and Unit-Identification Process**

We start with the bid ID, which is introduced in BNetzA's auction results. It is important to note that a bid may comprise several projects, all of which would share a bid ID and most characterisation data.

---

[7] As each TSO publishes this data separately, we combine four different payment registers into one dataset.
[8] The dataset contains data for all technology groups (solar, onshore wind, offshore wind and dispatchable RES)—but this study focuses on solar.

However, since projects within a single bid may differ in their size, location, and commissioning date, our analysis requires the introduction of an additional project ID[9] (referred to with subscript $i$, with $i \in [1, ..., n]$). This dataset also defines the auction of which the project was a part (denoted by superscript $a$) and the project's developer (denoted by superscript $d$).

The project's initial location, $loc_i^{in}$, is provided by the five-digit postal code of the planned construction site. The date of the initial public results announcement[10] specifies the starting date, $date_i^{in}$, as the date on which winning are told of their award. $date_i^{dline}$ refers to the date of the first deadline (i.e., the deadline without a tariff reduction). According to §54, art. 1 of the RES Act 2017, the first deadline is met 18 months after the final public announcement.[11] Furthermore, we can gather data on the total amount of capacity bid per auction, $bc^a$ (measured in MW), the tendered capacity, $tc^a$ (i.e., the amount of capacity BNetzA plans to award in the auction), the actual awarded capacity, $ac^a$, and the maximum awarded bid per auction, $mbid^a$.

Mapping with the second database (step 1 in Figure 2) is possible because the unit register includes the bid ID $i$ for all units that were part of an auction. Thus, bid IDs are searched for in the unit register to identify the auctioned projects that have been commissioned. For these projects, it is also possible to identify the size $cap_i$, the commissioning date $date_i^{end}$, the final project location $loc_i^{end}$ and a project-specific ID ($unit\_id_i$[12]). Projects not found in the unit register are assumed to be cancelled or unfinished. Note that, with this first step, it is already possible to gather or calculate most of the data required for the hypothesis testing (i.e., project duration, applicable penalties and developer metrics). The variables are listed in Table 3.

Due to the existence of multi-project and recurrent bidders, we aggregated the developers to calculate the developer-associated variables. A common practice among German bidders is to create a new energy company for each of—or a few of—their projects. While these companies generally bear different names, they are often registered under the same office (Grashof, 2019; Tews, 2018; Tiedemann et al., 2019). Therefore, the number of effective developers was measured by aggregating winning bidders whose company location was registered under the same address. This aggregation process led to a decrease in the total number of developers. Furthermore, for each developer, we defined a set, $J^d$, of won and constructed projects.

---

[9] The project ID is based on the bid ID. For example, project SOL17-2/048-3 is the third identified project pertaining to bid SOL17-2/048.
[10] The regulation presses BNetzA to publish the results as soon as possible; however, no time limit is established (§35, art. 3, RES Act 2017). On average, it takes BNetzA 30 days to publish results—but this timeframe can range from four days to six months. For AU1 to AU8, the first public announcement occurred around 15 days after the auction was held.
[11] The final public announcement occurs one week after the first announcement (§35, art. 2, RES Act 2017).
[12] The unit ID is a 33-digit code assigned to each commissioned RES unit.

**Table 3: Variables Resulting from BNetzA's Database-Matching Procedure (Step 1 in Figure 2)**

| Variable | definition | formulation | |
|---|---|---|---|
| $Dur_i$ | Difference between the project's commissioning and starting date | $Dur_i = date_i^{end} - date_i^{in}$ | (1) |
| $Pen_i^{dline}$ | Binary variable measuring whether the project faces a penalty due to late realisation | $Pen_i^{dline} = \begin{cases} 1 & if\ date_i^{end} > date_i^{dline} \\ 0 & Otherwise \end{cases}$ | (2) |
| $Pen_i^{loc}$ | Binary variable measuring whether the project faces a penalty due to location change | $Pen_i^{loc} = \begin{cases} 1 & if\ loc_i^{end} \neq loc_i^{in} \\ 0 & Otherwise \end{cases}$ | (3) |
| $Reg_i$ | Binary variable indicating whether the project's final location corresponds to a northern or a southern state; northern state $N$={Mecklenburg-Western Pomerania, Saxony-Anhalt, Schleswig-Holstein, Brandenburg, North-Rhine Westphalia, Lower Saxony} and southern state $S$={Baden-Wuerttemberg, Bavaria, Hesse, Rhineland-Palatinate, Thuringia, Saarland, Saxony}.[13] | $Reg_i = \begin{cases} 1 & if\ loc_i^{end} \in S \\ 0 & Otherwise \end{cases}$ | (4) |
| $Exp_i^{d,a}$ | Binary variable measuring the developer's level of experience. Takes the value of 0 if the developer has not won a bid before the awarding of project $i$ and the value of 1 if it has won at least one bid in a prior auction. In eq. 5, $A^b$ denotes the set of awarded projects before auction $a$ takes place.[14] | $Exp_i^{d,a} = \begin{cases} 1 & if\ \exists i: i \in J^d \cap A^b : b < a \\ 0 & Otherwise \end{cases}$ | (5) |
| $Size_i^d$ | Binary variable measuring the developer's size as the accumulated capacity won and realised by developer $d$ for the total program.[15] | $Size_i^d = \sum_{i=1}^{n} cap_i \quad \forall i \in J^d$ | (6) |

The data from BNetzA's auction results and unit register are, however, insufficient to identify the individual bid values. Hence, further information from the TSOs is needed.

Having matched bid IDs and unit IDs in the first step, we can use the unit IDs to match with BNetzA's unit register and the TSOs' payment register. This allows us to find historical data on generation and payments (step 2). Payments reported in the database for each unit include monthly market premium payments[16] and monthly/yearly side payments, such as avoided network charges, tariff reductions and bonuses. Once the payment information is identified, it is processed and cleaned before calculating the bid values (step 3). First, we use the tariff register to characterise each payment. Only payments under the regular market premium model are selected (i.e., all other side payments are removed; step 3.1). Second, the effective monthly market payments per kWh are calculated by dividing the gross payments by the associated generations. Due to the nature of the German auction system, this effective paid tariff does not correspond to what the bidders bid in the auction. Instead, it corresponds to the market premium, $mp$. To find the bids, the applicable monthly market value, $mv$ (i.e., the monthly average price paid for solar energy in the German spot market) is added (step 3.2). Finally, we calculated the net bid values, $BV_i^{net}$, using eq. (7) (step 3.3): Where $mp_{i,m}$ is the market premium paid by the TSO to unit/project $i$ in month $m$, and $mv_m$ is the market value in month $m$.

---

[13] The three remaining German states—the densely populated states of Berlin, Bremen and Hamburg—do not have winning or realised projects. Hence, all of the projects in our sample are in the northern or southern region.
[14] Note that experience is zero for all projects awarded in the first auction, $Exp_i^{d,1} = 0$.
[15] Note that size does not vary from auction to auction; rather, it is calculated for the whole program.
[16] Note that market premium payments vary by month. Hence, the payment register constitutes a much larger set of information than the unit register.

$$BV_{i,m}^{net} = mp_{i,m} + mv_m \qquad (7)$$

At the end of step 3, there is a net bid value for each month in which the unit received a payment. However, these values are still not the real bids made by developers. Since some projects are subject to tariff reductions due to delays or location changes, the corrected full bid values, $BV_i^{full}$, are calculated using eq. (8).

$$BV_{i,m}^{full} = BV_{i,m} + 0.3 pen_i^{loc} + 0.3 pen_i^{dline} \qquad (8)$$

Note that both $BV^{full}$ and $BV^{net}$ should be time invariant (i.e., constant over all months for each unit). Furthermore, all projects under the same bid should share the same full bid value, $BV^{full}$, but not necessarily the same net bid value, $BV^{net}$.[17] Moreover, since the second and third auctions—AU2 and AU3—were uniform-priced, all of the projects from these two auctions should have the same $BV^{full}$.[18]

Our dataset covers all available data for the German PV auction programme from AU1 in April 2015 to AU23 in October 2020. However, at the time of writing,[19] only the first twelve auctions have met their final construction deadlines, meaning that we can only properly analyse this set of auctions (AU1–AU12). Moreover, due to the delay[20] between BNetzA and TSO data, bid values can only be retrieved for projects with unit IDs built before 2020. As Table 4 shows, the data quality for bid values declines drastically from AU9 onwards. Therefore, the set of data from AU1 to AU8 is used for all analyses related to bid values.

Additionally, bid values cannot be estimated for all AU1–AU8 projects because (i) not all projects are realised, (ii) matching between databases is impossible due to inconsistencies in the reporting of unit IDs, which cannot be treated[21] and (iii), if the market price is higher than the bid value, the market premium is zero, meaning a bid value cannot be estimated. Point (iii) applies especially to projects commissioned in late 2018 and 2019, as increased market values coincide with lower bid values.

As shown in Table 4, of the 334 constructed projects reported by BNetzA for AU1 to AU8, 272 were assigned a unit ID (81%). Of those, 266 had a match in the TSOs' payment register—but a reliable bid value was only found for 232. In practice, since all projects within a bid should have the same bid value, it was possible to assign an observed full bid value, $BV_i^{full}$, to 13 additional projects even when a net bid value, $BV_i^{net}$, was not available. Hence, a total of 245 projects (73.4%) were completely identified. Finally, projects from AU2 and AU3 (57) were excluded from estimations of the bid values, as they do not represent the real bids; rather, they represent the value of the maximum awarded bid.

---

[17] This is due to the fact that not all projects covered by a single bid face the same tariff reduction.
[18] 8.49 €ct/kWh and 8 €ct/kWh, respectively.
[19] December 2020.
[20] At the moment, TSO data is usually updated once per year during the month of August.
[21] In contrast to metric data (e.g., a unit's payment for a particular month), missing values for unit characteristics cannot be estimated based on comparable observations.

As a result, a sample of 188 observations (245-57=188) from projects pertaining to AU1 and AU4–AU8 is used to answer questions concerning the bid values. Since the empirical complexity behind the calculation of project realisation and duration is less severe, a sample of 422 constructed projects from AU1 to AU12 is used to answer questions pertaining to these parameters.[22]

---

[22] Data validity and reliability are shown in Appendix D.

**Table 4: Data Availability per Auction**

| Auction | Date | Deadline | Awarded Capacity [MW] | No. of Awarded Bids | No. of Awarded Projects | No. of Built Projects | No. of Provided Unit IDs | TSOs Payments Found | Reliable Payment | Final Number of Bid Values |
|---|---|---|---|---|---|---|---|---|---|---|
| AU1 | Apr-15 | May-17 | 157 | 25 | 37 | 37 | 36 | 35 | 34 | 35 |
| AU2 | Aug-15 | Aug-17 | 159 | 33 | 48 | 41 | 35 | 34 | 24 | 25 |
| AU3 | Dec-15 | Dec-17 | 204 | 43 | 48 | 41 | 37 | 37 | 31 | 32 |
| AU4 | Apr-16 | Apr-18 | 128 | 21 | 30 | 30 | 29 | 28 | 25 | 28 |
| AU5 | Aug-16 | Aug-18 | 118 | 23 | 30 | 27 | 23 | 23 | 21 | 21 |
| AU6 | Dec-16 | Dec-18 | 163 | 27 | 50 | 49 | 36 | 36 | 30 | 33 |
| AU7 | Feb-17 | Feb-19 | 200 | 38 | 66 | 66 | 47 | 46 | 43 | 46 |
| AU8 | Jun-17 | Jun-19 | 201 | 32 | 42 | 42 | 29 | 27 | 24 | 25 |
| AU9 | Oct-17 | Oct-19 | 222 | 20 | 23 | 19 | 2 | 2 | 1 | 1 |
| AU10 | Feb-18 | Feb-20 | 201 | 24 | 31 | 21 | 10 | 6 | 4 | 6 |
| AU11 | Jun-18 | Jun-20 | 183 | 28 | 33 | 23 | 5 | 4 | 2 | 2 |
| AU12 | Oct-18 | Oct-20 | 192 | 37 | 48 | 26 | 7 | 2 | 1 | 1 |
| AU13 | Feb-19 | Feb-21 | 178 | 24 | 28 | 17 | 3 | | | |
| AU14 | Mar-19 | Apr-21 | 505 | 121 | 148 | 78 | 26 | 6 | 6 | 7 |
| AU15 | Jun-19 | Jun-21 | 205 | 14 | 26 | 6 | 2 | | | |
| AU16 | Oct-19 | Oct-21 | 153 | 27 | 33 | 7 | 4 | | | |
| AU17 | Dec-19 | Jan-22 | 501 | 121 | 137 | 31 | 8 | | | |
| AU18 | Feb-20 | Feb-22 | 101 | 18 | 18 | | | | | |
| AU19 | Mar-20 | Sep-22 | 301 | 51 | 54 | 3 | | | | |
| AU20 | Jun-20 | Sep-22 | 100 | 21 | 22 | | | | | |
| AU21 | Jul-20 | Sep-22 | 193 | 30 | 32 | 2 | 1 | | | |
| AU22 | Sep-20 | Oct-22 | 258 | 75 | 78 | | | | | |
| AU23 | Oct-20 | Nov-22 | 103 | 30 | 32 | | | | | |
| | AU1–AU8 | | 1330 | 242 | 352 | 334 | 272 | 266 | 232 | **245** |
| | AU1–AU12 | | 2129 | 351 | 486 | **422** | 295 | 280 | 240 | 254 |

Note that the realisation deadlines in column three have only been reached for AU1–AU12 at time of writing (December 2020). Hence, the table shows a low number of built projects. Exploring these incomplete auction data would not reveal stable results and, thus, we have excluded them from the following analysis.

## 3.3 Testing Hypotheses and Answering Questions

The methodology employed to answer the first four questions raised in Section 3.1 is straightforward, as it merely involves the analysis of the program averages and occurrence percentages described in Section 3.2. To answer the remaining three questions, however, we must introduce a hypothesis-testing framework. This section details our approach to solving each individual research question.

We address Q1 (on the realisation of projects) by using the realisation rate per auction, defined as the sum of the capacities of built projects, by total awarded capacity. Since each auction comprises a set of awarded projects, $A^a$, the realisation rate is calculated as:

$$RR^a = \frac{\sum_{i=1}^{n} cap_i}{ac^a} \quad \forall i \in A^a \tag{9}$$

To answer Q2 (on the time it takes German developers to construct their projects), we use the average duration per auction, $Dur^a$, defined as:

$$Dur^a = \frac{\sum_{i=1}^{n} Dur_i}{ac^a} \quad \forall i \in A^a \tag{10}$$

We also calculate the percentage of total projects built after the first deadline by dividing the sum of the projects built late (i.e., $Pen_i^{dline} = 1$) by the number of total built projects. The percentage for each auction is reached as follows:

$$Bl^a = \frac{\sum_{i=1}^{n} Pen_i^{dline}}{\sum_{i=1}^{n} Status_i} \quad \forall i \in A^a \tag{11}$$

To answer Q3 (on project location), we find the share of projects with a location change per auction as follows:

$$Lchg^a = \frac{\sum_{i=1}^{n} Pen_i^{loc}}{\sum_{i=1}^{n} Status_i} \quad \forall i \in A^a \tag{12}$$

We address Q4 (on the difference between the net and full bid values) by comparing the auction averages for the sample of the net and full bid values, $BV_i^{net}$ and $BV_i^{full}$. Note that these measures can all be aggregated over AU1–AU8 to find the programme's average.

To answer Q5 (on the relationship between geographical aspects, bids and project duration), we test for the significance of the difference in means between groups. For the regional distribution, $Reg_i$, we compare the average duration ($Dur_i$) and the average bid value ($BV^{full}$) of projects built in the north and those built in the south. The occurrence in location change, $Pen_i^{loc}$, is analysed using the marginal bid, $Bmg$, instead of the average bid value as a proxy for the subsidy level. This exchange is done due to the negative and significant correlation between time and both bid values and location change. Since our variables are not normally distributed, hypotheses $H0_{5.1-5.4}$ are tested using a Mann–Whitney–Wilcoxon test.

Q6 asks whether the auction design favours large players. We have defined this advantage in three categories. The first category explores barriers to auction entry; $H0_{6.1}$ tests for entry barriers for new developers while $H0_{6.2}$ tests for entry barriers for small developers. New developers are defined as those who have not won a bid in a previous auction; for each project, we measure whether a project is a part of the first bid won by developer $d$. Small developers are defined as those who accumulate 2 MW[23] or less of realised capacity over the course of the programme.

$$New_i^{d,a} = \begin{cases} 1 & if \ \nexists i: i \in \{J^d \cap A^b: b < a\} \\ 0 & Otherwise \end{cases} \quad (13)$$

$$Small_i^{d,a} = \begin{cases} 1 & if \ Size_i^d \leq 2 \\ 0 & Otherwise \end{cases} \quad (14)$$

The sum of new and small developers per auction is then used to assess $H0_{6.1}$ ('The proportion of new developers decreases steadily over time') and $H0_{6.2}$ ('The proportion of small developers among awarded bids decreases over time'). While a pronounced decline is expected in the first few rounds due to the accumulation of 'old' developers, a stabilisation of the share of new and small developers in later auctions would indicate that they can still be competitive.

The second category evaluates whether large developers have an advantage to win higher profits by bidding closer to the marginal bid. We evaluate $H0_{6.3}$ by testing the significance of the correlation between developer size, $Size_i^d$, and the difference from the maximum awarded bid, $Bmg_i^{d,a}$, where:

$$Bmg_i^{d,a} = mbid^a - BV_i^{full} \quad (15)$$

A negative correlation between the variables would indicate that large developers are better at estimating the marginal bid, meaning they have an advantage to realise higher profits.

The third category evaluates whether lack of experience results in underbidding and project delays by testing the relationship between experience, $Exp_i^{d,a}$, and both bids, $BV_i^{full}$, and project duration, $Dur_i$. $H0_{6.4}$ and $H0_{6.5}$ evaluate the significance of the difference in means between the groups using a Mann–Whitney–Wilcoxon test. Significance here would support the hypothesis that projects constructed by inexperienced developers differ structurally in terms of their duration and necessary subsidy level.

Answering Q7 requires the definition of two additional measures: level of competition and technology costs. We measure competition in each auction using the bid-to-cover ratio, $Bcr_i^a$, defined as:

---

[23] This value was arbitrarily set, as there is no established definition of 'small developer' in the literature. However, we tested for various levels between 1 and 5 MW and produced similar results.

$$Bcr_i^a = \frac{bc^a}{tc^a} \tag{16}$$

We believe that $Bcr_a$ is a good measure of competition, as excess demand reduces the exercise of market power by buyers. As a proxy for technology costs, we use the PV price index, which is available six months after the auction date.[24] The use of the 6–month–lag PV price index ($Pvc6_{+i}^a$) entails the assumption that developers purchase modules six months after being awarded a project in an auction, which aligns with developers' expectation of decreasing costs.

$H0_{7.1}$ and $H0_{7.2}$ are assessed by testing the significance of their coefficients resulting from the following regression:

$$BV_i^{full} = \beta_1 + \beta_2 N\_Pvc6_{+i}^a + \beta_3 N\_Bcr_i^a \tag{17}$$

Note that we use the normalised values of $Pvc6_{+i}^a$ and $Bcr_i^a$ to obtain comparable coefficients along the same scale.

## 4 RESULTS AND DISCUSSION

### 4.1 Realisation Rates

We find high realisation rates with several auctions reaching near-full-capacity deployment, as depicted in Figure 3. Between AU1 and AU12, the programme's average $RR$ reaches 82%. Consequently, the German auction programme can be considered as effective, as the capacity built per year has always reached the targeted 400 MW (FFAV, 2015, para. 1). The overall high level of realisation is especially surprising considering the low penalties for non-compliance. Developers face a maximum penalty of 50 €/kW for unfulfilled awarded capacity, meaning that, after two years, they only risk about 7% of their total project costs.[25]

---

[24] This figure is based on the monthly PV price index published by pvXchange. The metric is calculated as the average PV price index for the different regions (Germany, Japan/Korea, China and South-East Asia/Taiwan) from January 2015 to July 2017 and the average PV price index for the different module types (High Efficiency, All Black, Mainstream and Low Cost) from August 2017 onwards. We chose the values applicable six months after the auctions, as they provided the highest correlation with the bid values.

[25] Calculation based on average installation costs of 700 €/kW for large-scale PV installations (Fraunhofer ISE, 2018).

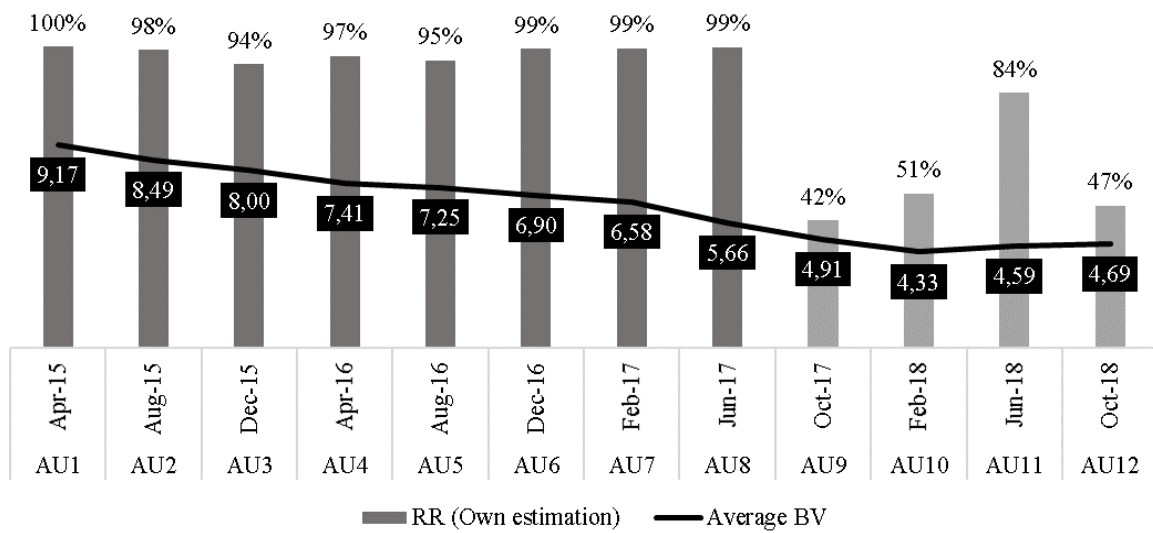

**Figure 3: Capacity Realisation Rates (Own Estimation)**

Looking closely at Figure 3, the difference in realisation rates between AU1–AU8 and AU9–AU12 is stunning. While the first group has an average realisation rate of 97%,[26] the second group has one of only 56%. Our working hypothesis to explain this striking difference is that PV module prices were falling faster than expected after the first eight auctions, making investments for the successful bidders attractive and, in turn, leading to high realisation rates despite a continuous decline in auction bids. The low realisation rate for AU9–AU12 coincides not only with the lowest observed subsidy levels but also with relatively stable PV module costs (see Figure 4). Hence, any successful bidder in AU9–AU12 who calculated bids based on the expectation of continuously falling PV module costs underestimated those costs. The policy implications are clear: If this hypothesis is correct, the effectiveness of this programme is a transitory phenomenon anchored to developments in technology costs.

---

[26] Our result is in line with the 96.5% reported by BNetzA for the same period, validating our data.

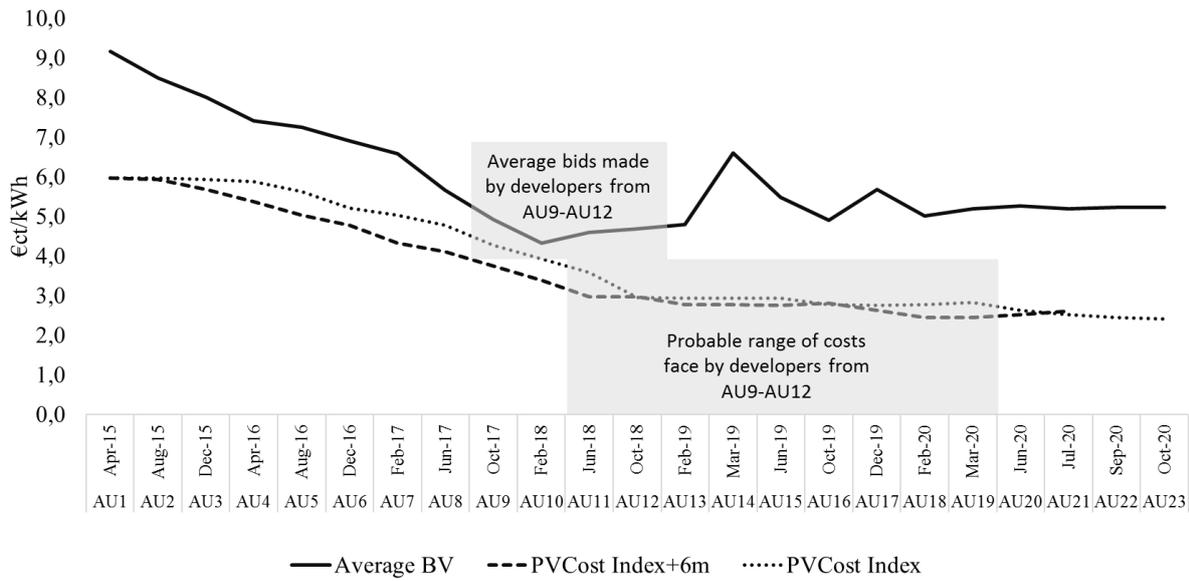

**Figure 4: Average Awarded Bid vs. PV Price Index as a Proxy for Developer Costs**

## 4.2 Average Project Duration and Share of Projects Built after the First Deadline

Based on the completed projects from AU1–AU12, developers take an average of 542 days (about 1.5 years) to complete their projects. Weighting by capacity (rather than projects), the average duration changes only slightly to 527 days. The average construction time has remained fairly constant over time (see Figure 5).

The fact that PV capacity can consistently be brought online after only about 1.5 years is a policy-relevant result. In a country like Germany, where the construction of conventional power plants take 5–10 years (between the final investment decision and the beginning of commercial operations), a combination of PV and storage can be reliably and effectively used to provide the Germany system with additional electrical energy.

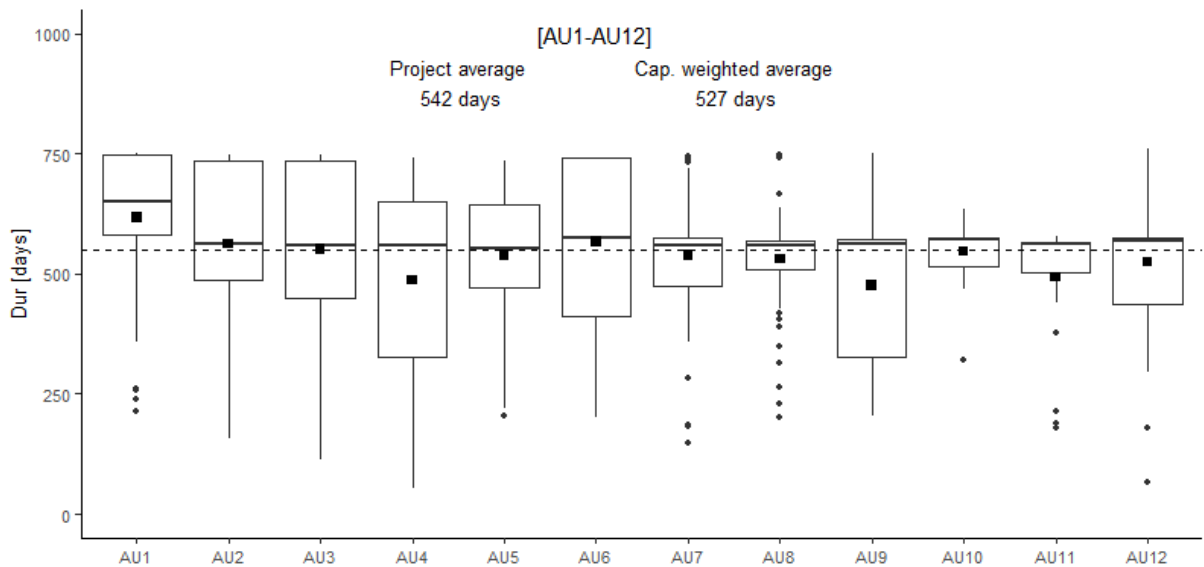

**Figure 5: Average Project Duration**

Another important observation is the high share of projects realised within 18 months of the winners being announced, meaning that most projects were 'on time' and received full subsidies. As shown in Figure 6, the majority of projects (72% of realised projects and 78% of realised capacities) are finished before the first deadline. However, this implies that 28% of realised projects (22% of realised capacities) are completed after the first deadline and incur a tariff reduction of 0.3 €/kWh (see Table 5). Late realisation can be caused by two reasons:

a) The project could not be finished earlier, and finishing late is more profitable than not finishing at all.
b) The project could have been finished earlier, but the developer intentionally waited. This could be driven by decreases in PV costs outweighing the tariff reduction. In fact, 0.3 €/kWh represents just 3–7% of the average auction bid, while the prices of PV modules declined by an average of 10% between the first and final construction deadlines.

The second explanation would align with the steep increase in realisation rates shortly before the deadlines; if falling module prices are driving the timeline, investors would wait as long as possible—but finish just before one of the two deadlines. This hypothesis is also backed by the narrow distribution of project duration around the average in the most recent auctions (see Figure 5). This behaviour suggests that developers have learned how to better manage their projects and build without penalties; alternatively, it could simply mean that developers are more pressured to speed up construction just before the deadlines, when the likelihood of a significant decline in costs becomes less probable.

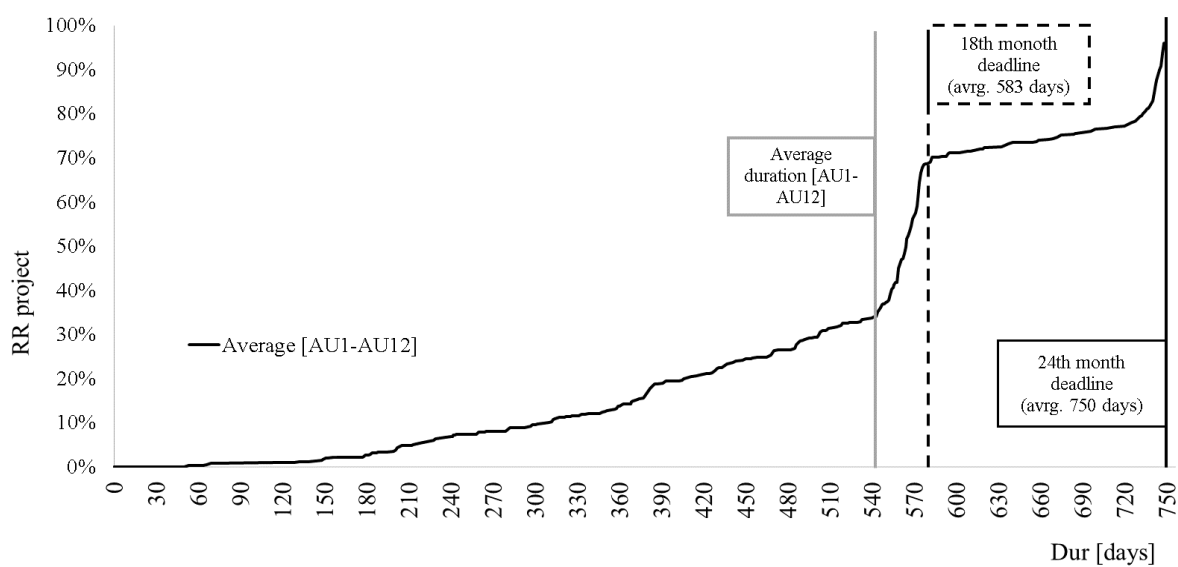

**Figure 6: Project Realisation Rate Against Project Duration**

## 4.3 Share of Projects with Location Change

The share of projects from AU1–AU12 that changed locations amounts to 46% (33% in terms of capacity)—though with a downwards trend (see Table 5). Location changes were exceptionally common (more than two-thirds) for the first two auctions—possibly due to the novelty of the auction process. Overall, the high location-change rates show that developers find the flexibility of the auction design to be useful—despite the incurred tariff reduction of 0.3 €/kWh. This flexibility also increases realisation rates, as investors prefer to realise projects at a new location with a lower payment instead of forfeiting a successful bid and paying the non-compliance penalty.

**Table 5: Percentage of Projects Subject to Tariff Reductions**

| Auction | Date | $Lchg$[27] | | $Bl$[28] | |
|---|---|---|---|---|---|
| | | Project | Capacity | Project | Capacity |
| AU1 | Apr-15 | 68% | 56% | 54% | 36% |
| AU2 | Aug-15 | 68% | 57% | 39% | 22% |
| AU3 | Dec-15 | 32% | 32% | 34% | 33% |
| AU4 | Apr-16 | 27% | 7% | 27% | 25% |
| AU5 | Aug-16 | 48% | 39% | 30% | 30% |
| AU6 | Dec-16 | 53% | 26% | 37% | 26% |
| AU7 | Feb-17 | 58% | 35% | 23% | 24% |
| AU8 | Jun-17 | 40% | 33% | 19% | 13% |
| AU9 | Oct-17 | 21% | 18% | 11% | 2% |
| AU10 | Feb-18 | 43% | 49% | 19% | 14% |
| AU11 | Jun-18 | 30% | 23% | 4% | 0% |
| AU12 | Oct-18 | 27% | 4% | 23% | 40% |
| AU1–AU8 | | 50% | 36% | 32% | 26% |
| AU1–AU12 | | 46% | 33% | 27% | 16% |

---

[27] As defined in eq. (12)(12).
[28] As defined in eq. (11).

In terms of distribution, Figure 7 shows that built projects broadly follow Germany's solar irradiation patterns, with most projects built in Bavaria. However, this south-to-north trend is not absolute. A negative exception is Baden-Württemberg, with very low participation relative to its level of irradiation.[29] Furthermore, both Brandenburg and Mecklenburg-Vorpommern have high capacity shares despite moderate solar irradiation. Land availability and policy support for renewable invest are the two main drivers of these disparities. In Germany, the construction of subsidised, large-scale PV is only allowed in conversion areas or 'less-favoured farming areas'. Bavaria has introduced less-restrictive construction rules, falling into the second category. Therefore, developers are incentivised to construct projects in Bavaria rather than in its neighbour state Baden-Württemberg, as it is easier—and likely cheaper—to find suitable locations (IWR, 2019).

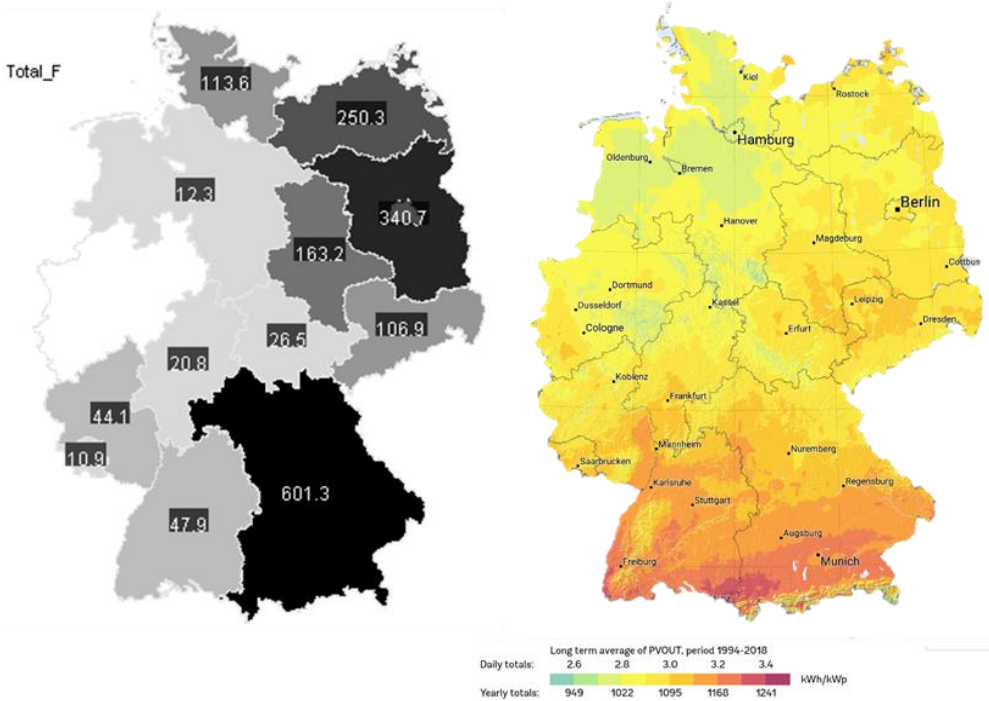

**Figure 7: Regional Distribution in MW for AU1–AU12 and PV Power Potential in Germany (World Bank & Solargis, 2019)**

As described earlier, nearly one-third of all projects change location between bid and realisation despite the associated penalty. Our data enable the analysis of underlying trends regarding the location movements (e.g., from one federal state to another). Of the 195 projects that changed location (46% of all AU1–AU12 constructed projects), the majority—around 60%—moved to another federal state.

Bavaria and Brandenburg are the two states with the highest planned and realised capacities. Hence, it is unsurprising that most changes occur within these states. However, as shown in Figure 8, there is a surprising, significant degree of relocation between the two. While Bavaria benefits from 59 MW of net additions, Brandenburg faces net losses of 60.6 MW. We are unsure of the reason behind this interesting fact; it could certainly be addressed in future research. Other states are awarded less capacity to begin with and tend to lose even more towards realisation.

---

[29] This phenomenon is not specifically related to the auction format. While 25% of Germany's total installed PV capacity is of a large-scale nature, this figure is only 8% in Baden-Württemberg (Ministry of Environment, Climate and Energy, Baden-Württemberg, 2019).

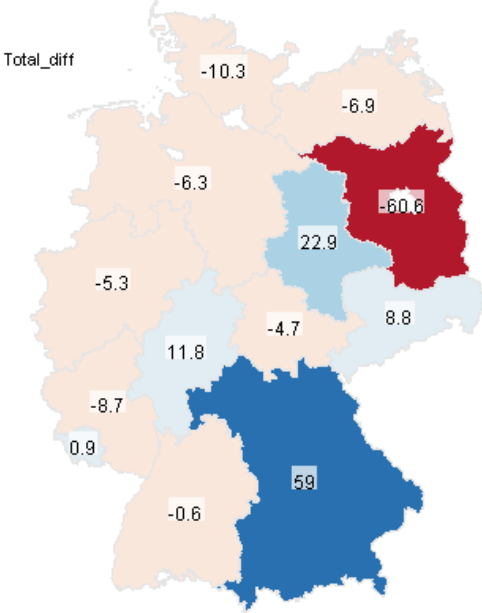

**Figure 8: Location Change by State AU1–AU8**

## 4.4 Penalties and Net Bid Values

The number of bids facing a penalty is significant. Table 5 shows that 46% of projects change location and, thus, face a payment reduction. Additionally, while Section 4.2 revealed that the average project is completed before the 18-month deadline, a non-negligible share of AU1–AU12 projects (28% of projects and 22% of capacity) are realised between the 18- and 24-month deadlines, meaning they face a penalty for delayed construction. Moreover, many projects face both penalties, as location change is often associated with delayed construction. These penalties translate into a reduction of full bid values.

Net bid values (i.e., bid values corrected for penalties) are interesting, as they provide further insights into the competitiveness of PV. If we interpret average net bids as proxies for LCOEs,[30] it is meaningful that projects are realised despite net bids being significantly lower than full bid values.

As shown in **Figure 9: Full Bid Values vs Net Bid Values per Auction and on Average**, due to penalties, realised AU1–AU8 projects receive, on average, a subsidy that is 0.2 €/kWh lower than their original bids. Hence, PV in Germany is even more competitive than published auction results suggest. **Figure 9: Full Bid Values vs Net Bid Values per Auction and on Average** also shows that marginal bidders are not subject to penalties, as the maximal bid is the same for full and net bid values. This behaviour is expected, as securing higher subsidy levels makes investors less likely to delay construction in the hopes of decreasing technology costs.

---

[30] For an exact fit, several assumptions would need to be met, including perfect competition and *ex-ante* rational expectations of penalty likelihood.

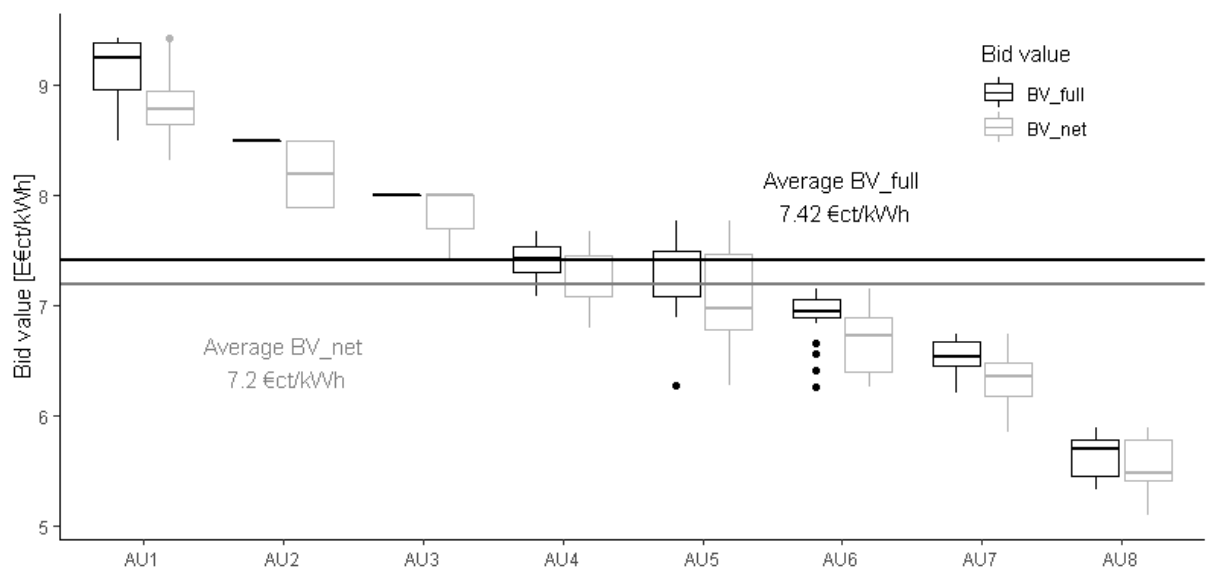

**Figure 9: Full Bid Values vs Net Bid Values per Auction and on Average**

## 4.5 Do Geographical Aspects Affect Bids and Construction Periods?

To answer this question, we evaluate whether projects changing location ($Pen_i^{loc} = 1$) demonstrate average construction times and required subsidy levels that are distinct from those among projects built at the original reported construction site (i.e., the subset with $Pen_i^{loc} = 0$). In the same way, we use $Reg_i$ to measure whether projects built in northern states structurally differ from those built in southern states. These questions are formalised by hypotheses $H0_{5.1-4}$.

- $H0_{5.1}$: There is no difference in average project duration between projects with and without location change
- $H0_{5.2}$: There is no difference in average subsidy level between projects with and without location change
- $H0_{5.3}$: There is no difference in average project duration between projects built in the north and those built in the south of Germany
- $H0_{5.4}$: There is no difference in average subsidy level between projects built in the north and those built in the south of Germany

Our results are presented in Figure 10: Effect of Region and Location Change on Duration and Bid Values (Two-Sided Mann–Whitney–Wilcoxon Test)Figure 10. Note that for $H0_{5.1}$, the subsidy level is measured using the difference from the marginal bid, $Bmg$, instead of the bid values due to the positive correlation that time has with both $BV$ and $Pen_i^{loc}$. Using $BV$ would have certainly biased the results.

Regarding $H0_{5.1}$, we found a significant difference between the groups in terms of average construction time. Therefore, $H0_{5.1}$ can be rejected. We find that bids with a location change are completed in an average of 598 days (20 months) while bids without a location change are completed in an average of 495 days (16.5 months). Thus, the difference between the two groups is 103 days (3.4 months).

The significantly longer average construction time for projects with a change of location also implies a high correlation between delayed realisation and location change.[31] Given the penalty for location change, it would seem reasonable for bidders to reveal the planned location when participating in an auction. There are two main reasons for location changes. First, a location change could mean there were unforeseen problems with realisation at the original location. For example, an investor could plan for a certain site when bidding, but unforeseen legal difficulties could force them to find a new site. Second, investors could confidently bid on a project without having a planned location, as they are certain that they have the resources to realise it within the specified timeframe. Both reasons would generally increase a project's duration, especially relative to a smoothly running project with a pre-specified location).

Regarding $H0_{5.2}$, the Mann–Whitney–Wilcoxon test demonstrates that a significant difference between the groups cannot be established; hence, $H0_{5.2}$ cannot be rejected. Our results support the idea that developers are likely not changing location strategically; rather, they are forced to find new sites due to unforeseeable problems. Otherwise, projects with location changes would, on average, receive higher subsidy levels.

In terms of regional differences, $H0_{5.3}$ can be rejected, as we find that projects in the south can be realised significantly faster. Projects in the south take, on average, 33 fewer days to be built than projects in the north. Finally, $H0_{5.4}$ cannot be rejected. It is particularly interesting that we found no difference between regions in terms of subsidy levels. One potential explanation is that higher land costs in the south compensate for its higher degree of solar irradiation.

---

[31] Note that the penalties are additive, meaning that a location change and a delay result in a sizable total tariff reduction of 0.6 €ct/kWh.

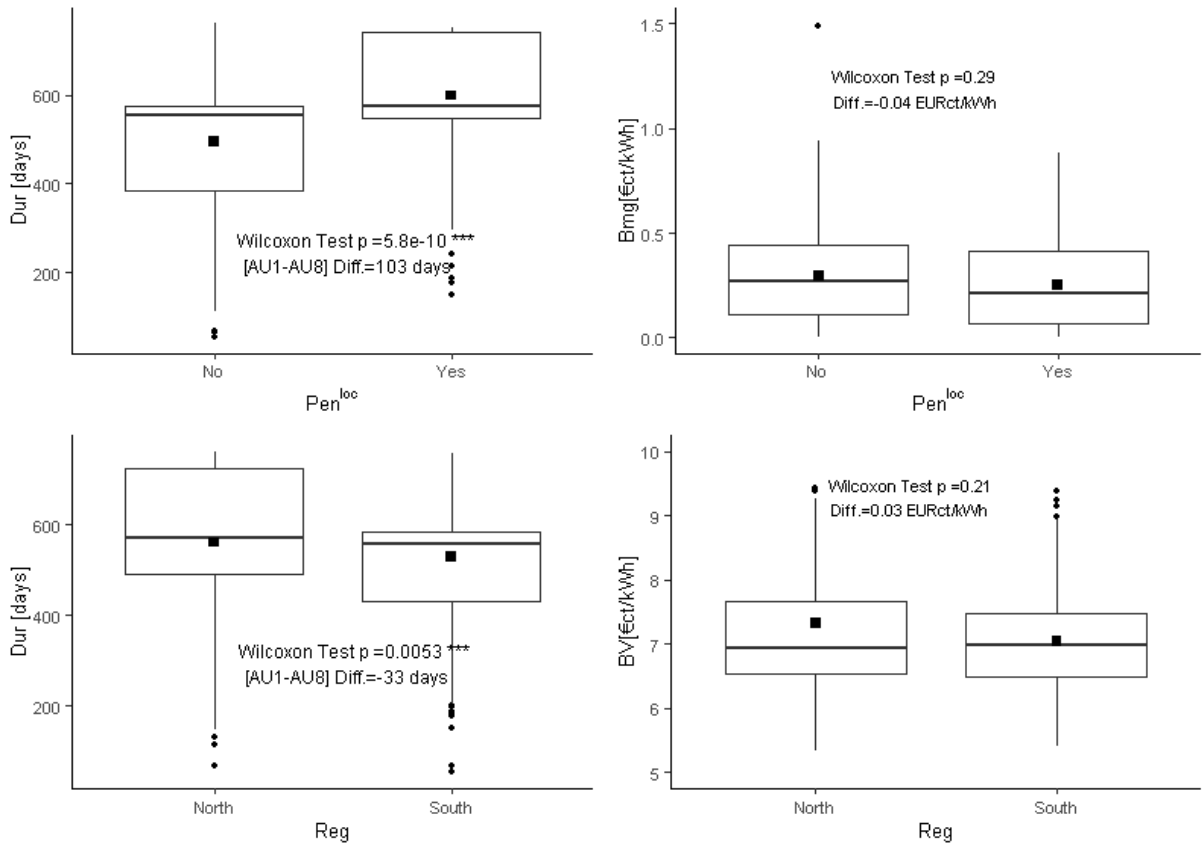

Note: *** 1% sig. level; ** 5% sig. level.

**Figure 10: Effect of Region and Location Change on Duration and Bid Values (Two-Sided Mann–Whitney–Wilcoxon Test)**

## 4.6 Does the Auction Design Favour Large Players?

We address three different ways in which the auction design may favour large players. First, we evaluate whether the design incorporates strong entry barriers for inexperienced or small developers. Second, we evaluate whether large developers have an advantage estimating the marginal bid and, in turn, winning disproportionally high profits. Third, we evaluate whether inexperienced developers are prone to underbidding and construction delays. These three approaches are formalised by hypotheses $H0_{6.1-5}$.

- $H0_{6.1}$: The proportion of new developers decreases steadily over time
- $H0_{6.2}$: The proportion of small developers among awarded bids decreases over time
- $H0_{6.3}$: There is no correlation between developer size (capacity) and accuracy at estimating the marginal bid
- $H0_{6.4}$: There is no difference in average project duration between experienced and inexperienced developers
- $H0_{6.5}$: There is no difference in average subsidy level between projects realised by experienced developers and those realised by inexperienced developers

### 4.6.1 Are Inexperienced and Small Developers Being Pushed Out of the Market?

In terms of new developers,[32] the auction programme shows a steep decline between AU1 and AU4. However, this result was unavoidable, as all bidders were—by definition—'new' in AU1. As Figure 11 shows, the proportion of new developers from AU4 to AU12, while certainly volatile, does not show a significant trend in either direction. The result is similar for small[33] developers; their share is volatile (and, in fact, reaches zero for AU1 and AU7) but fluctuates around 15% over time. There is no systematic decline for either measure. Thus, while large and experienced developers constitute a larger share, hypotheses $H0_{6.1}$ and $H0_{6.2}$ can be rejected.

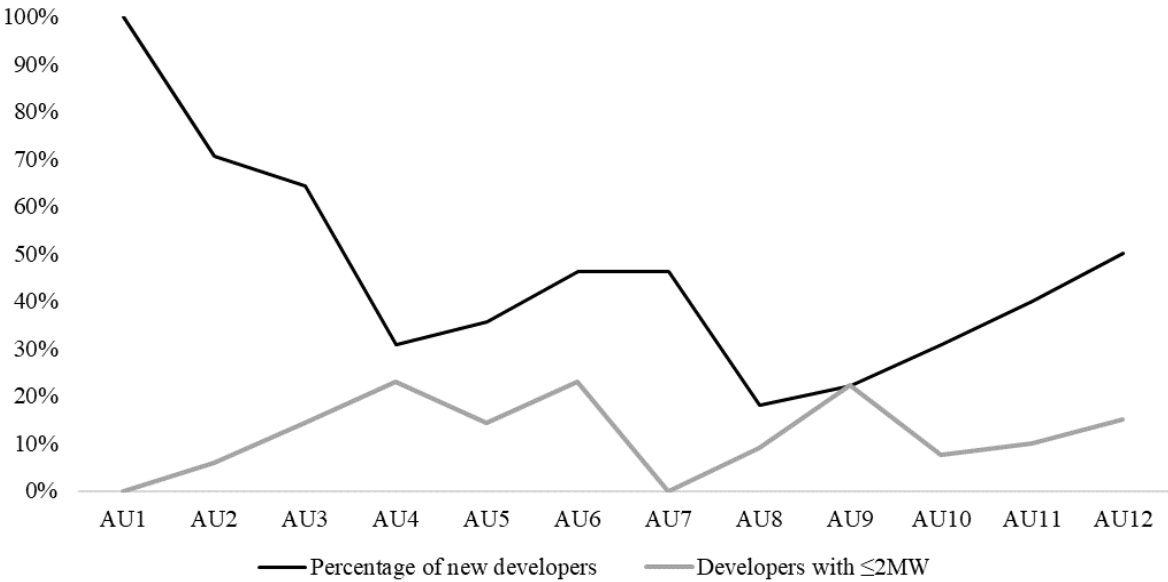

**Figure 11: Participation Rate of New and Small Developers**

### 4.6.2 Do Large Developers Have an Advantage to Win Higher Profits?

To answer $H0_{6.3}$, we analysed the individual bid values' distance from the marginal bid and the degree of correlation between this distance and developer size. Our correlation test shows a negative and significant correlation of -0.18; in other words, bids from larger developers are generally closer to the marginal bid). Consequently, we reject $H0_{6.3}$. This result implies higher profits for larger developers, giving them a competitive advantage against smaller participants. This finding is in line with the theory, and it makes sense that larger developers would be better at reading the market. However, as shown in Figure 12, larger developers also commonly misjudge the maximum awarded bid, with several large bubbles landing at the bottom.

---

[32] New developers were defined in eq. 14 as those who have not won a bid in previous auctions.
[33] Small developers were defined in eq. 15 as those with 2 MW or less of accumulated capacity over the course of the auction programme.

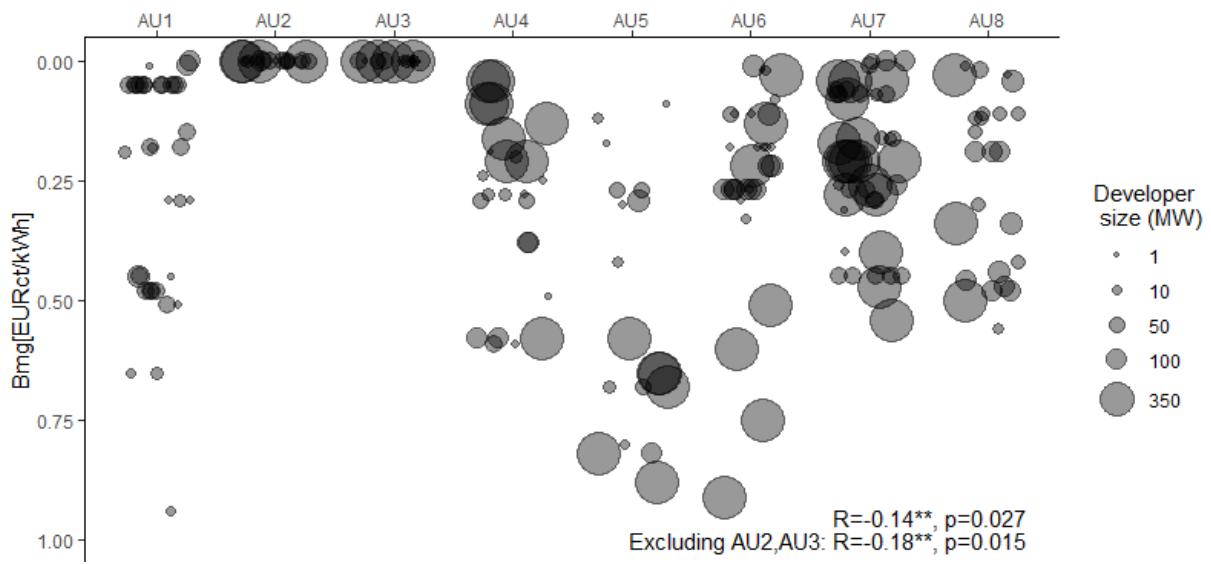

Note: The lower the difference, the closer to the marginal bid. *** 1% sig. level; ** 5% sig. level.
**Figure 12: Difference from the Marginal Bid by Developer Size**

### 4.6.3 Does a Lack of Experience Result in Project Delays or Underbidding?

Finally, we assess $H0_{6.4}$ and $H0_{6.5}$ by testing the difference between experienced and inexperienced developers in average project duration and subsidy level. As shown in Figure 13, our results suggest that experienced developers are not significantly better than inexperienced developers at improving their profits; they do not place significantly higher bids near the marginal or avoid project delays to a significantly greater degree. Consequently, hypotheses $H0_{6.4}$ and $H0_{6.5}$ can be rejected.

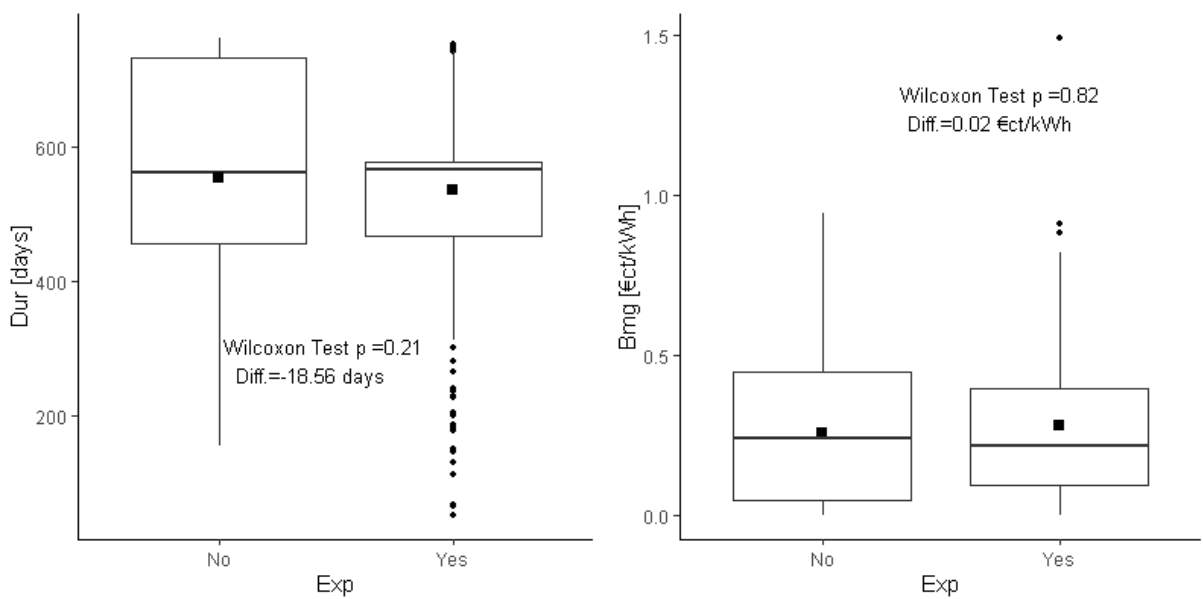

Note: Difference in group means is not significant. *** 1% sig. level; ** 5% sig. level.
**Figure 13: Effect of Experience on Bid Values and Project Duration (Two-Sided Mann–Whitney–Wilcoxon Test)**

Overall, this section supports the idea that larger developers have advantages over smaller and inexperienced developers. However, the auction design does not necessarily reinforce these advantages. While our results are limited by our sample which does not include unawarded bids, our findings are in line with the Italian case from Cassetta et al. (2017) and the South African case from Kruger et al. (2021). Cassetta et al. (2017) regressed the bid values of both awarded and unawarded projects from a wind-onshore programme on a set of variables including project size and developer experience; finding no significant effect of these last two variables on bid values. Kruger et al. (2021) used the Herfindahl–Hirschman Index to analyse market concentration in both PV and wind auctions, concluding that concentration is present to a small degree but mostly due to developers' advantages outside the auction design. It is important to highlight the fact that our results do not contest the idea that large developers benefit from economies of scale and experience. Rather, they indicate that developers are not incorporating additional benefits into their bids (i.e., they are not increasing their chances of being awarded by placing lower bids).

## 4.7 Does Competition Influence Bid Values?

The effect of competition and its relative importance compared to technology cost reductions is assessed through the following hypotheses.

- $H0_{7.1}$: Competition has a significant negative effect on bid values
- $H0_{7.2}$: The reduction in technology costs has been a major factor behind the auction results

We find that competition[34] ($N\_Bcr$) significantly influences bid values—the higher the $N\_Bcr$, the lower the bids. However, we also find that the effect of PV module costs is significantly higher. Therefore, $H0_{7.1}$ and $H0_{7.2}$ cannot be rejected. These results (details in Table 6) are derived from the econometric model established in eq. (17). Our model has an R-squared of 95%, and both coefficients are significant at α=1%.

**Table 6: Econometric Model**

| | Estimate | Std. Error | t-value | Pr(>|t|) | Sig. |
|---|---|---|---|---|---|
| Intercept | 6.13 | 0.03 | 216.65 | 0.00 | *** |
| $N\_Pvc6_+$ | 1.70 | 0.04 | 45.97 | 0.00 | *** |
| $N\_Bcr$ | -0.24 | 0.02 | -14.69 | 0.00 | *** |
| Residual standard error | 0.247 on 182 df | | | | |
| n | 185 | | | | |
| Multiple R-squared | 0.95 | | | | |
| Adjusted R-squared | 0.949 | | | | |
| F-statistic | 1727 *** | | | | |

Note: *** 1% sig. level; ** 5% sig. level.

---

[34] Measured as an auction's bid-to-cover ratio, defined in eq. (16).

For an average scenario (i.e., a scenario in which panel costs and competition level are equal to their mean values), the subsidy would be 6.14 €/kWh. A decrease in technology costs by one standard deviation would decrease the subsidy level by 1.70 €/kWh. This is in line with our expectations—lower procurement costs decrease bids. Furthermore, a decrease in competition by one standard deviation would increase bids by 0.24 €/kWh. While this effect is significantly smaller, the direction is also as expected. To sum up, both reduced panel costs and increased competition have contributed to low bids in PV auctions—but panel costs have had a more pronounced impact.

# 5 CONCLUSION AND POLICY IMPLICATIONS

This paper presented a methodology to analyse the results of the German renewable auction data at an unprecedented level of detail. In particular, we identified individual projects' parameters, such as location, subsidy level and several others that were not detailed in this article. Our approach can be used to analyse all RES auctions in Germany, including onshore and offshore wind. In this article, we used our approach to conduct an in-depth analysis of the German solar auction programme.

We found that the programme effectively deployed solar generation capacities. The results of the first 12 auctions demonstrate an average realisation rate of 82%. Therefore, the net target for deployment through auctions (400 MW per year) has been achieved. However, the level of realisation has varied greatly over time. AU1–AU8 had an average realisation rate of 97% while AU9–AU12 had an average realisation rate of only 56%. PV module prices decreased significantly over the course of the first eight auctions, enabling low-cost project realisation; from AU9 to AU12, however, PV module prices were relatively stable. The expectations of lower module prices following these auctions did not materialise, making bids that expected these declines unfeasible. Based on these results, we concluded that the extremely high realisation rates for AU1–AU8 and the very low realisation rates for AU9–AU12 constitute a transitory phenomenon. Future realisation rates will be highly dependent on developments in PV module prices.

Furthermore, we found that the average project duration is 1.5 years. Most projects are completed within 18 months and, thus, face no penalty for late realisation. However, 28% of realised projects are completed past the 18-month deadline, leading to reduced net payments. Additionally, we found that 46% of the commissioned projects changed location. Interestingly, we found a correlation between duration and location change: projects that changed location took, on average, 20% longer (3.4 additional months) than those built at their planned site. Observing that more than a quarter of projects finish late and that nearly half of all projects change location, our results indicate that project developers have made extended used of the auction programme's design flexibility, realising their projects despite penalties.

Looking at the regional distribution, auctioned projects are not structurally different from those financed under FIT. In both cases, the distribution mainly follows Germany's solar irradiation patterns and, to a lesser extent, Germany's land regulations.

In contrast to existing studies on auction evaluation (Bayer et al., 2018b; Cassetta et al., 2017; del Río et al., 2015; del Río and Mir-Artigues, 2019; Gephart et al., 2017; Hochberg and Poudineh, 2018; IRENA, 2017; Müsgens and Riepin, 2018), our data suggest that project duration and bid values are not strongly affected by developer size and experience.

Finally, there is no doubt that the decrease in PV module costs has been the major driving factor behind the marked reduction in auction bids. However, under the German auction design, heavy competition leads developers to adapt their bid structures to changes in their costs; thus, auctions seem to potentialize the effect of falling PV module costs. Moreover, competition could play a more significant role when PV module costs are stable or increasing.

**Table 7: Summary of Results**

| Question | Result |
|---|---|
| Are projects being realised? | Yes. On average, 82% of capacity is realised. This number is 97% for the first 8 auctions and 56% for AU9 to AU12. We discuss the reasons behind this decline in the results section |
| How long do German developers take to construct PV projects? | 1.5 years |
| What percentage of projects change location? | 46% |
| What is the difference between the net bid values (i.e., including penalties for delays and location changes) and the full bid values (i.e., the actual bids)? | 0.2 €/kWh |
| Do geographical aspects affect bids and construction periods? | |
| $H0_{5.1}$: There is no difference in average project duration between projects with and without location change | Reject. The construction of projects with a location change take, on average, 3.4 months longer |
| $H0_{5.2}$: There is no difference in average subsidy level between projects with and without location change | Accept |
| $H0_{5.3}$: There is no difference in average project duration between projects built in the north and those built in the south of Germany | Reject. The construction of projects built in the north take, on average, one month longer |
| $H0_{5.4}$: There is no difference in average subsidy level between projects built in the north and those built in the south of Germany | Accept |
| Does the auction design favour large players? | |
| *Smaller developers are pushed out of the market* | Reject |
| $H0_{6.1}$: The proportion of new developers decreases steadily over time | Reject. The proportion of new developers fluctuates consistently around 40% |
| $H0_{6.2}$: The proportion of small developers among awarded bids decreases over time | Reject. The proportion of small developers consistently fluctuates around 15% |
| *Large developers have an advantage to win higher profits* | Accept |
| $H0_{6.3}$: There is no correlation between developer size (capacity) and accuracy at estimating the marginal bid | Reject |
| *A lack of experience may result in underbidding or project delays* | Reject |
| $H0_{6.4}$: There is no difference in average project duration between experienced and inexperienced developers | Accept |

| Question | Result |
|---|---|
| $H0_{6.5}$: There is no difference in average subsidy level between projects realised by experienced developers and those realised by inexperienced developers | Accept |
| Does competition influence bid values? | Accept |
| $H0_{7.1}$: Competition has a significant negative effect on bid values | Accept |
| $H0_{7.2}$: The reduction in technology costs has been a major factor behind the auction results | Accept |

# REFERENCES

Anatolitis, V., Welisch, M., 2017. Putting renewable energy auctions into action – An agent-based model of onshore wind power auctions in Germany. Energy Policy 110, 394–402. https://doi.org/10.1016/j.enpol.2017.08.024

Batz, T., Müsgens, F., 2019. A first analysis of the photovoltaic auction program in Germany. Presented at the International Conference on the European Energy Market, EEM. https://doi.org/10.1109/EEM.2019.8916472

Bayer, B., Berthold, L., Moreno Rodrigo de Freitas, B., 2018a. The Brazilian experience with auctions for wind power: An assessment of project delays and potential mitigation measures. Energy Policy 122, 97–117. https://doi.org/10.1016/j.enpol.2018.07.004

Bayer, B., Schäuble, D., Ferrari, M., 2018b. International experiences with tender procedures for renewable energy – A comparison of current developments in Brazil, France, Italy and South Africa. Renewable and Sustainable Energy Reviews 95, 305–327. https://doi.org/10.1016/j.rser.2018.06.066

Black, A., 2005. Performance Based Incentives for PV via auction. Presented at the Proceedings of the Solar World Congress 2005: Bringing Water to the World, Including Proceedings of 34th ASES Annual Conference and Proceedings of 30th National Passive Solar Conference, pp. 2260–2263.

Cassetta, E., Monarca, U., Nava, C.R., Meleo, L., 2017. Is the answer blowin' in the wind (auctions)? An assessment of the Italian support scheme. Energy Policy 110, 662–674. https://doi.org/10.1016/j.enpol.2017.08.055

del Río, P., 2017. Designing auctions for renewable electricity support. Best practices from around the world. Energy for Sustainable Development 41, 1–13. https://doi.org/10.1016/j.esd.2017.05.006

del Río, P., Haufe, M.-C., Wigand, F., Steinhilber, S., 2015. Overview of Design Elements for RES-E Auctions.

del Río, P., Mir-Artigues, P., 2019. Designing auctions for concentrating solar power. Energy for Sustainable Development 48, 67–81. https://doi.org/10.1016/j.esd.2018.10.005

Eberhard, A., Kåberger, T., 2016. Renewable energy auctions in South Africa outshine feed-in tariffs. Energy Science & Engineering 4, 190–193. https://doi.org/10.1002/ese3.118

Fraunhofer Ise, 2018. Levelized Cost of Electricity- Renewable Energy Technologies.

Fraunhofer Ise, Agora Energiewende, Consentec GmbH, TU Wien | Energy Economics Group, 2014. Auctions for Renewable Energy in the European Union: Questions Requiring further Clarification.

Freiflächenausschreibungsverordnung – FFAV, 2015.

Gephart, M., Klessmann, C., Wigand, F., 2017. Renewable energy auctions – When are they (cost-)effective? Energy and Environment 28, 145–165. https://doi.org/10.1177/0958305X16688811

Grashof, K., 2019. Are auctions likely to deter community wind projects? And would this be problematic? Energy Policy 125, 20–32. https://doi.org/10.1016/j.enpol.2018.10.010

Hochberg, M., Poudineh, R., 2018. Renewable auction design in theory and practice: lessons from the experience of Brazil and Mexico. Oxford Institute for Energy Studies. https://doi.org/10.26889/9781784671068

IEA, 2020. Renewables 2020 - Analysis and forecast to 2025.

IRENA, 2018. Renewable power generation costs in 2017.

IRENA, 2017. Renewable Energy Auctions: Analysing 2016.

IRENA, 2013. Renewable Energy Auctions in Developing Countries.

Kreiss, J., Ehrhart, K.-M., Haufe, M.-C., 2017. Appropriate design of auctions for renewable energy support – Prequalifications and penalties. Energy Policy 101, 512–520. https://doi.org/10.1016/j.enpol.2016.11.007

Kruger, W., Eberhard, A., 2018. Renewable energy auctions in sub-Saharan Africa: Comparing the South African, Ugandan, and Zambian Programs. Wiley Interdisciplinary Reviews: Energy and Environment 7, e295. https://doi.org/10.1002/wene.295

Kruger, W., Nygaard, I., Kitzing, L., 2021. Counteracting market concentration in renewable energy auctions: Lessons learned from South Africa. Energy Policy 148, 111995. https://doi.org/10.1016/j.enpol.2020.111995

Mora, D., Islam, M., Soysal, E.R., Kitzing, L., Blanco, A.L.A., Förster, S., Tiedemann, S., Wigand, F., 2017. Experiences with auctions for renewable energy support, in: 2017 14th International Conference on the European Energy Market (EEM). Presented at the 2017 14th International Conference on the European Energy Market (EEM), pp. 1–6. https://doi.org/10.1109/EEM.2017.7981922

Müsgens, F., Riepin, I., 2018. Is offshore already competitive? Analyzing German offshore wind auctions. Presented at the International Conference on the European Energy Market, EEM. https://doi.org/10.1109/EEM.2018.8469851

REN21, 2018. Renewables 2018 Global Status Report.

Sach, T., Lotz, B., von Blücher, F., 2018. Auctions for the support of renewable energy in Germany.

Shrimali, G., Konda, C., Farooquee, A.A., 2016. Designing renewable energy auctions for India: Managing risks to maximize deployment and cost-effectiveness. Renewable Energy 97, 656–670. https://doi.org/10.1016/j.renene.2016.05.079

Tews, K., 2018. The crash of a policy pilot to legally define community energy. Evidence from the German auction scheme. Sustainability (Switzerland) 10. https://doi.org/10.3390/su10103397

Tiedemann, S., Bons, M., Sach, T., Jakob, M., Klessmann, C., Anatolitis, V., Billerbeck, A., Winkler, J., Höfling, H., Kelm, T., Metzger, J., Jachmann, H., Bangert, L., Maurer, C., Tersteegen, B., Hirth, L., Reimann, J., Ehrhart, K.-M., Hanke, A.-K., Navigant Energy Germany GmbH, 2019. Evaluierungsbericht der Ausschreibungen für erneuerbare Energien.

Tiedemann, S., Förster, S., Wigand, F., 2016. Auctions for Renewable Energy Support in Italy: Instruments and Lessons Learnt. Aures.

Toke, D., 2015. Renewable Energy Auctions and Tenders: How good are they? International Journal of Sustainable Energy Planning and Management 8, 43–56. https://doi.org/10.5278/ijsepm.2015.8.5

USAID, 2016. Technical Note: The Basics of Competition & Auctions for Renewable Energy.

Voss, A., Madlener, R., 2017. Auction schemes, bidding strategies and the cost-optimal level of promoting renewable electricity in Germany. Energy Journal 38, 229–264. https://doi.org/10.5547/01956574.38.SI1.avos

Welisch, M., Kreiss, J., 2019. Uncovering Bidder Behaviour in the German PV Auction Pilot: Insights from Agent-based Modeling. The Energy Journal 40. https://doi.org/10.5547/01956574.40.6.mwel

Winkler, J., Magosch, M., Ragwitz, M., 2018. Effectiveness and efficiency of auctions for supporting renewable electricity – What can we learn from recent experiences? Renewable Energy 119, 473–489. https://doi.org/10.1016/j.renene.2017.09.071

# APPENDIX

## Appendix A: German PV Auction Design and Regulation

| Topic of Regulation | Provisions | RES Act Section |
|---|---|---|
| German Federal Network Agency (BNetzA) | Institution in charge of conducting and monitoring RES auctions. | §85 |
| Volume | See Appendix B. | §28, art. 2 |
| Maximum value | Must be adapted based on §49 regarding the reduction of general PV tariffs. | §37b |
| Project construction deadline | Without tariff reduction: 18 months after the public announcement of the award. With tariff reduction: 24 months after the public announcement of the award. | §54, art. 1 and §37d, art. 2 |
| Project location | The bids must include the location in which the project will be built, with details on the land, district, municipality, local sub-district and cadastral parcel. For ground-mounted installations, the bid must be supplemented by the bidder's declaration that they are the owner of the site on which the solar installation is to be erected or are submitting the bid with the approval of the site's owner. | §30, art. 1, num. 6 and § 37, art. 2 |
| Bid size | The minimum bid size allowed is 750 kW. | §30, art. 2 |
| Multi-bid | Bidders may submit several bids for different installations. | §30, art. 3 |
| Multi-project | A bid may comprise various projects. | §54, art. 1 |
| Securities | The security that must be paid amounts to 50 €/kW. An initial security of 5 €/kW must be paid when the bid is placed; a second security of 45 €/kW must be paid by the tenth working day following the public announcement of the award. The second payment can be reduced to 20 €/kW if additional land-use documentation is provided alongside the bid. | §31 and §37a |
| Announcement of the awards | BNetzA is obliged to report the following information: Bid deadline, total capacity awarded, bidder names, location, bid ID, lowest and highest bid values, capacity-weighted average bid. | §35 |
| Expiry of awards | Bids expire if the second security is not paid or if the projects are not commissioned within 24 months of the public announcement of the award. | §37d, art. 2 |
| Tariff reduction | There is a reduction of 0.3 €/kWh if the units are not commissioned within 18 months of the winner being notified. If a bid has been subdivided into several projects, the reduction applies only to the projects that have been delayed. There is a reduction of 0.3 €/kWh if the final construction location does not, at least in part, conform to the location indicated in the bid. | §54, art. 1–2 |
| Penalties | If the second security is not paid, the penalty is equal to the first security (i.e., 5 €/kW). If 5% or more of the bid is cancelled (not built within the 24-month period), the penalty amounts to 50 €/kW of the cancelled capacity. The tariff can be reduced to 25 €/kW if the second security was reduced through the provision of additional land-use documentation. | § 55, art. 3 |

# Appendix B: Tender Schedule (Capacities in MW)

| Year | Total | Feb | Mar* | Apr | Jun | Jul* | Aug | Sep* | Oct | Dec* |
|---|---|---|---|---|---|---|---|---|---|---|
| 2015 | 1150 | | | 800 | | | 150 | | | 200 |
| 2016 | 410 | | | 125 | | | 125 | | | 160 |
| 2017 | 600 | 200 | | | 200 | | | | 200 | |
| 2018 | 600 | 200 | | | 200 | | | | 200 | |
| 2019 | 1475 | 175 | 500 | | 150 | | | | 150 | 500 |
| 2020 | 1800 | 100 | 300 | | 150 | 300 | | 400 | 150 | 400 |
| 2021 | 1950 | 150 | 400 | | 100 | 400 | | 400 | 100 | 400 |
| 2022 ff. | 600 | 200 | | | 200 | | | | 200 | |

Note: *Special tenders taking place in 2019–2021. Source: FFAV (2015), §3, art. 1 and §28, art. 2; RES ACT (2017), §2a.

If the tendered capacity for one year is not completely awarded, the difference is added to the next year's tendered capacity and distributed equally among the tender dates. Capacities are also reduced by three factors: 1) the sum of solar capacity installed in Germany but won in auctions in other EU countries during the previous year; 2) the sum of large-scale installed solar capacity registered during the previous year that was not part of an auction programme; 3) half of the solar capacity installed during the previous year that was won in the neutral wind-solar auction programme.

# Appendix C: Summary of Nomenclature and Equations

| Directly Extracted Information | | | |
|---|---|---|---|
| Parameter | Description | Parameter | Description |
| $i$ | Awarded projects | $m$ | Month |
| $a$ | Auction | $A^a$ | Set of projects awarded in auction $a$ |
| $d$ | Developers | $J^d$ | Set of projects won by developer $d$ |
| $mp_{i,m}$ | Market premium paid by the TSOs | $date_i^{in}$ | Final announcement of winners |
| $mv_m$ | Market value | $date_i^{dline}$ | Project's first construction deadline |
| $loc_i^{in}$ | Initial reported location | $date_i^{end}$ | Project's commissioning date |
| $loc_i^{end}$ | Actual construction site | $cap_i$ | Project capacity |
| $mbid^a$ | Maximum awarded bid | $bc^a$ | Total bid capacity per auction |
| $pvc6_{+i}^a$ | PV price index six months after the auction[35] | $tc^a$ | Tendered capacity per auction |
| Indirectly Extracted Information | | | |
| Variable | Description and Related Question | | Calculation |
| $Status_i$ | Project status | Q1 | $Status_i = \begin{cases} 1 & if\ \exists\ date_i^{end} \\ 0 & Otherwise \end{cases}$ |

---

[35] This figure is based on the monthly PV price index published by pvXchange. The metric is calculated as the average PV price index for the different regions (Germany, Japan/Korea, China and South-East Asia/Taiwan) from January 2015 to July 2017 and the average PV price index for the different module types (High Efficiency, All Black, Mainstream and Low Cost) from August 2017 onwards. We chose the values applicable six months after the auctions, as they provided the highest correlation with the bid values.

| | | | |
|---|---|---|---|
| $Dur_i$ | Project duration | Q2 | $Dur_i = date_i^{end} - date_i^{in}$ |
| $Pen_i^{dline}$ | Penalty due to late realisation | Q4 | $Pen_i^{dline} = \begin{cases} 1 & if\ date_i^{end} > date_i^{dline} \\ 0 & Otherwise \end{cases}$ |
| $Pen_i^{loc}$ | Penalty for change of location | Q4 | $Pen_i^{loc} = \begin{cases} 1 & if\ loc_i^{end} \neq Loc_i^{in} \\ 0 & Otherwise \end{cases}$ |
| $Reg_i$ | Project's regional location (north/south) | Q5 | $Reg_i = \begin{cases} 1 & if\ loc_i^{end} \in S \\ 0 & Otherwise \end{cases}$ |
| $Size_i^d$ | Developer's size | Q6 | $Size_i^d = \sum_{i=1}^{n} cap_i \quad \forall i \in J^d$ |
| $Exp_i^{d,a}$ | Developer's experience | Q6 | $Exp_i^{d,a} = \begin{cases} 1 & if\ \exists i: i \in \{J^d \cap A^b: b < a\} \\ 0 & Otherwise \end{cases}$ |
| $BV_{i,m}^{net}$ | Project's net bid value | Q4 | $BV_{i,m}^{net} = MP_{i,m} + MV_m$ |
| $BV_{i,m}^{full}$ | Project's full bid value | Q4 | $BV_{i,m}^{full} = BV_{i,m} + 0.3Pen_i^{loc} + 0.3Pen_i^{dline}$ |
| $RR^a$ | Realisation rate per auction | Q1 | $RR^a = \frac{\sum_{i=1}^{n} cap_i}{ac^a} \quad \forall i \in A^a$ |
| $Dur^a$ | Average duration per auction | Q2 | $Dur^a = \frac{\sum_{i=1}^{n} Dur_i}{ac^a} \quad \forall i \in A^a$ |
| $Bl^a$ | Percentage of projects built after the first deadline | Q2 | $Bl^a = \frac{\sum_{i=1}^{n} Pen_i^{dline}}{\sum_{i=1}^{n} Status_i} \quad \forall i \in A^a$ |
| $Lchg^a$ | Percentage of projects with location change | Q3 | $Lchg^a = \frac{\sum_{i=1}^{n} Pen_i^{loc}}{\sum_{i=1}^{n} Status_i} \quad \forall i \in A^a$ |
| $New_i^{d,a}$ | Developer with no experience prior to the time of the bid for project $i$ | Q6 | $New_i^{d,a} = \begin{cases} 1 & if\ \nexists i: i \in \{J^d \cap A^b: b < a\} \\ 0 & Otherwise \end{cases}$ |
| $Small_i^{d,a}$ | Developer is small if they accumulate 2 MW or less of realised capacity over the course of the programme. | Q6 | $Small_i^{d,a} = \begin{cases} 1 & if\ Size_i^d \leq 2 \\ 0 & Otherwise \end{cases}$ |
| $Bmg_i^{d,a}$ | Difference from the maximum awarded bid | Q6 | $Bmg_i^{d,a} = mbid^a - BV_i^{full}$ |
| $Bcr_i^a$ | Bid-to-cover ratio as a proxy for the competition level | Q7 | $Bcr_i^a = \frac{bc^a}{tc^a}$ |
| | Model | Q7 | $BV_i^{full} = \beta_1 + \beta_2 pvc6_{+i}^{\ a} + \beta_3 Bcr_i^a$ |

## Appendix D: Data Validation

The complexity of the data-gathering and sample-selection processes requires the validation of observations. We checked the validity of our sample and the reliability of our bid-reconstruction process by comparing the estimated capacity-weighted average bid of our sample with the one reported by BNetzA in the auction results. As depicted in the figure below, the averages are nearly identical. Furthermore, all observations for AU2 and AU3 match the values reported by BNetzA (8.49 €/kWh and 8 €/kWh, respectively).

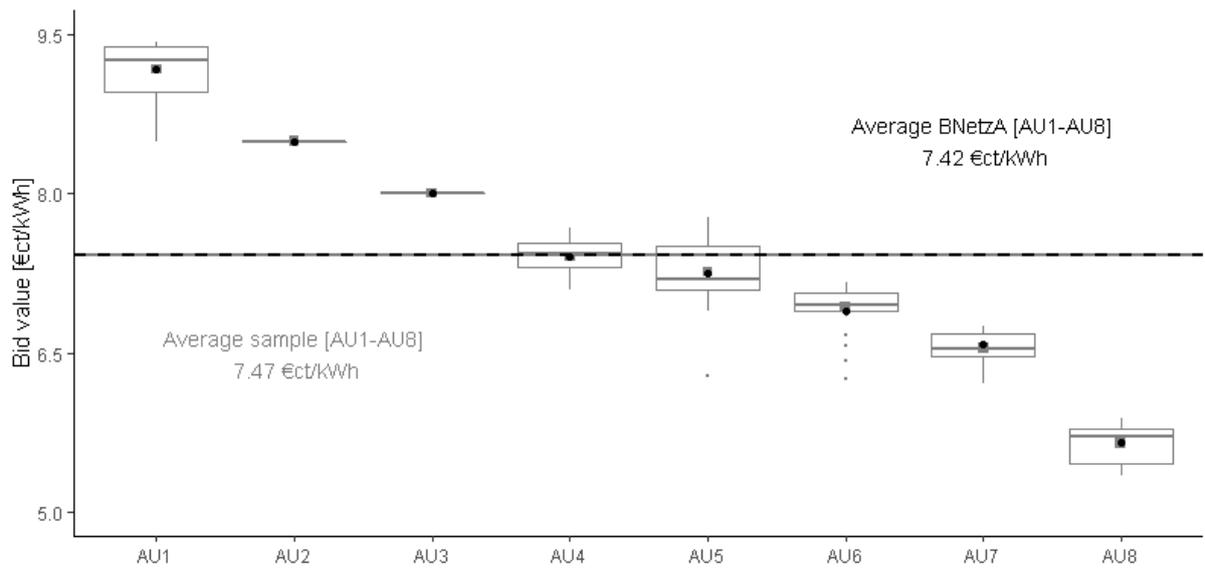

**Reported vs. Estimated Capacity-Weighted Average Bid**